\documentclass[11pt]{article}
\usepackage{graphicx, amsmath, color, xcolor} % Required for inserting images
\usepackage{comment}
\usepackage{bm}
\usepackage{subfiles}
\usepackage{amsthm}
\usepackage{multirow, array, booktabs, threeparttable}
\usepackage{hyperref}
\newtheorem{theorem}{Proposition}

\DeclareSymbolFont{matha}{OML}{txmi}{m}{it}% txfonts
\DeclareMathSymbol{\varv}{\mathord}{matha}{118}

\graphicspath{{images/}}
\usepackage[margin=1in]{geometry}
\usepackage{setspace}

\usepackage[backend=biber,
            sorting = none, 
            style = nejm, 
            autocite = superscript, %change to inline to view
            url = false]{biblatex}
\usepackage[noperiod]{jabbrv}

\DeclareFieldFormat{journaltitle}{\mkbibemph{#1}}
\DeclareFieldFormat{number}{\mkbibparens{#1}}

\renewbibmacro*{journal+issuetitle}{%
  \usebibmacro{journal}%
  \setunit{\addperiod\addspace}% % Added \addperiod here
  \usebibmacro{date}%
  \setunit{\addsemicolon}%
  \printfield{volume}%
}

\title{Target trial emulation without matching: a more efficient approach for evaluating vaccine effectiveness using observational data}

\author{Emily Wu, Elizabeth Rogawski McQuade, Mats Stensrud, Razieh Nabi,  David Benkeser}

\begin{document}
%\onehalfspacing
\maketitle

\section*{Abstract}
Real-world vaccine effectiveness has increasingly been studied using matching-based approaches, particularly in observational cohort studies following the target trial emulation framework. Although matching is appealing in its simplicity, it suffers important limitations in terms of clarity of the target estimand and the efficiency or precision with which is it estimated. Scientifically justified causal estimands of vaccine effectiveness may be difficult to define owing to the fact that vaccine uptake varies over calendar time when infection dynamics may also be rapidly changing. We propose a causal estimand of vaccine effectiveness that summarizes vaccine effectiveness over calendar time, similar to how vaccine efficacy is summarized in a randomized controlled trial. We describe the identification of our estimand, including its underlying assumptions, and propose simple-to-implement estimators based on two hazard regression models. We apply our proposed estimator in simulations and in a study to assess the effectiveness of the Pfizer-BioNTech COVID-19 vaccine to prevent infections with SARS-CoV2 in children 5-11 years old.  In both settings, we find that our proposed estimator yields similar scientific inferences while providing significant efficiency gains over commonly used matching-based estimators.

\section*{Keywords}

Vaccine effectiveness; Causal inference; Estimands; Cohort design; COVID-19; Target trial emulation; Matching

\newpage

\section{Introduction}

While randomized controlled trials (RCTs) are the gold standard for establishing vaccine efficacy, studies of vaccines in real-world settings are also critical for public health decision-making. Real-world vaccine effectiveness (VE) has increasingly been studied using target trial emulation \autocite{hernan2016_specifying, hernan2016_using, komura2025_exploring}.
Target trial emulation implies that the start of follow-up in an observational vaccine study should be the time an eligible individual receives a vaccine. However, it is challenging to define ``vaccination'' time for unvaccinated individuals. 
%A  challenge associated with observational vaccine studies is the lack of a well-defined start of follow-up time for a control comparator group.
A common solution is to match vaccinated to unvaccinated participants on a set of confounders and define the start of follow-up for each matched pair as the time of vaccination for the vaccinated individual
\autocite{dagannoa2021_bnt162b2, barda2021_effectiveness, reis2021_effectiveness, ioannou2022_covid19, cohen-stavichandraj.2022_bnt162b2, monge2023_effectiveness, hulme2023_challenges, meah2023_design}. An advantage of matching-based analyses is their simplicity \autocite{suissa2023_prevalent, didelez2024_re}; assuming the matching procedure adjusts sufficiently for confounding, a matched dataset can be analyzed using methods similar to those used in randomized trials. However, matching also has several important limitations. First, matching can change the target population, a feature often overlooked by researchers \autocite{greenland1990_matching, 
westreich2010_invited, shiba2021_using}. Second, if the set of matching variables is high-dimensional, it may be difficult to identify matches for all observations, thereby reducing the precision of the analysis \autocite{stuart2010_matching, hernan2020_causal}. Finally, matching is often performed without a clear articulation of the target estimand, making it difficult to interpret the output as a well-defined causal effect \autocite{rose2011_open, keogh2023_causal}. 
 
%Finally, matching is a procedure wherein the target estimand is defined by the procedure itself, rather than via explicit scientific reasoning, sometimes leading to a lack of clarity regarding interpretation and the assumptions required for appropriate interpretation .

We present an alternative to matching for observational vaccine effectiveness studies using target trial emulation \autocite{dang2023_start}. We explicitly describe a causal estimand for vaccine effectiveness and the assumptions under which it is identified. Moreover, we provide simple-to-implement estimators that show dramatic efficiency improvements over matching estimators in practice.

%To begin, we provide relevant background information on randomized controlled trials of vaccines that are emulated by common matching procedures (Section 2.1). We then review the ``rolling-cohort" matching design used in previous studies and examine the corresponding matching-based estimand (Section 2.2). Based on these observations, we propose a general class of causal vaccine effectiveness estimands (Section 3.1), which shares similarities with the estimands in randomized clinical trials and matching studies. We further propose plug-in estimators for the proposed estimands that are based on two straightforward hazard-based regression models (Section 3.2). Simulations show that the proposed estimator produces point estimates similar to those of a matching-based estimator while enjoying significantly improved efficiency (Section 4).  We illustrate the method using a study of the effectiveness of the COVID-19 vaccine in children aged 5-11 years during the 2021-2022 school year (Section 5) and end with a discussion (Section 6). 

\section{Review of statistical estimands in vaccine studies}

Our causal estimand is motivated by the statistical estimands evaluated in randomized controlled trials of vaccines and the matching procedures commonly used to emulate them.  We review the design and estimands of these studies, focusing on estimands based on cumulative incidences \autocite{hernan2010_hazards}. We emphasize how the estimands account for (i) calendar time over which participants ``enroll'' in the study and (ii) covariate values for participants that are included in the analysis-- factors that will be important when we later seek to establish causal estimands quantifying vaccine effectiveness. For simplicity, we initially assume no right-censoring of study endpoints and later relax this assumption. Without loss of generality, we consider studies of single-dose vaccines. 

\subsection{Randomized controlled trials}

In most individually randomized vaccine efficacy trials, individuals are enrolled in the trial over a period of time. Let $D$ denote the day an individual enrolls in the trial relative to the start day of the trial. On the day of enrollment, participants are individually randomized to receive an active vaccine or a placebo/control vaccine, denoted by $V$. At enrollment, it is common to collect covariate information on participants, which could include demographics (e.g., age, sex), information derived from pre-vaccination serum samples (e.g., baseline serostatus), and other relevant medical information (e.g., co-morbidities). We denote this information by the vector $X$ and without loss of generality assume that $X$ assumes a finite number of values. 

After enrollment, participants are followed until the first instance of the primary endpoint -- typically symptomatic disease with infection confirmed by a diagnostic, although other endpoints may be considered \autocite{hudgens2004_endpoints, mehrotra2021clinical}. We use $T$ to denote the time from enrollment to first instance of the primary endpoint. A comparison of the cumulative incidence of the primary endpoint within $t_0$ days of vaccination,  $P(T \leq t_0 \mid V = v)$, between vaccine arms could be used to establish vaccine efficacy. However, it is common to exclude cases occurring during the $\tau$ days following vaccination when the vaccine-induced immune response may still be building \autocite{hofner2024_vaccine, hudgens2004_endpoints, dean2018_design, dean2019_design}. We define $R(\tau) = I(T > \tau)$ as an indicator of remaining endpoint-free $\tau$ days after vaccination. The modified marginal cumulative incidence is $P(T \leq t_0 \mid V = v, R(\tau) = 1)$ and a common target of inference is the cumulative vaccine efficacy, defined as $\text{VE}_{\text{RCT}}(t_0) = 1 - P(T \leq t_0 \mid V = 1, R(\tau) = 1) / P(T \leq t_0 \mid V = 0, R(\tau) = 1)$.

We note that, although $X$ and $D$ do not explicitly appear in the formulation of $\text{VE}_{\text{RCT}}(t_0)$, both may influence incidence of the primary endpoint. For example, $X$ may include risk factors for the primary endpoint such as age or geographical location, while $D$ may be associated with the incidence of the primary endpoint due to changing background transmission and/or emerging variants. To make the influence of these factors explicit, we can write the marginal cumulative incidence as a weighted average of the $D$- and $X$-specific cumulative incidence,  $P(T \leq t_0 \mid V = v,  R(\tau) = 1, D = d, X = x)$: 
\begin{equation}
\begin{aligned} 
&P(T \leq t_0 \mid V = v, R(\tau) = 1) \\
&\hspace{2em} = \sum_{x,d} \bigg [ P(T \leq t_0 \mid V = v,  R(\tau) = 1, D = d, X = x) \\
&\hspace{7em} \times P(D = d \mid V = v, R(\tau) = 1, X = x) P(X = x \mid V = v, R(\tau) = 1) \bigg] \ . \label{eqn:cuminc_rct}
\end{aligned}
\end{equation}

This representation highlights that the marginal cumulative incidences, and therefore $\text{VE}_{\text{RCT}}(t_0)$, typically estimated in randomized trials depend on the particular distribution of enrollment days and covariates observed in the trial. We note that, in randomized trials, assuming that there are minimal protective vaccine effects prior to day $\tau$, we expect the enrollment-day distributions ($P(D = d \mid V = v, R(\tau) = 1, X = x)$) and covariate distributions ($P(X = x \mid V = 1, R(\tau) = 1) $) to be similar across vaccine arms. Thus, while $D$ and $X$ may be associated with incidence of the primary endpoint, they are unlikely to be associated with vaccine status $V$ and therefore are not confounders. As a result, we expect that $\text{VE}_{\text{RCT}}(t_0)$ will be an unbiased measure of a trial-specific vaccine efficacy. As we move into observational settings, we will need to explicitly control for confounding of outcomes by $D$ and $X$.

\subsection{Observational cohort studies of vaccine effectiveness and matching} \label{sec:obs_studies_ve}

In an observational cohort study of vaccine effectiveness, individuals meeting  eligibility criteria on a pre-specified study start date are identified from a data source such as electronic medical records. Covariate information $X$ on these individuals can be derived, and we let $V^*$ be an indicator of whether an individual is vaccinated during the study period. We let $D^*$ denote the day an individual receives vaccination relative to study start; for unvaccinated individuals, $D^* = \infty$. We use $Y$ to denote the time from study start to first occurrence of the study endpoint. 

In contrast to a randomized trial, individuals who receive vaccination in an observational setting may differ from individuals who do not in terms of risk for the endpoint and calendar period of exposure \autocite{lipsitch2016_observational, hulme2023_challenges, fung2024_sources}. Thus, covariates $X$ and vaccination timing $D^*$ must be used to control for confounding. While regression modeling and sequential trial approaches have occasionally been used to this end \autocite{vasileiou2021_interim, lin2022_effectiveness, mcconeghy2022_infections, demonte2024_assessing}, rolling cohort-matching methods are far more popular, especially in target trial emulation studies \autocite{dagannoa2021_bnt162b2, pearce2023_are, komura2025_exploring}. 

In rolling-cohort designs, individuals vaccinated after the study start are identified. Newly vaccinated individuals on day $d$ are identified and matched with other eligible individuals who have similar covariate values but are unvaccinated on day $d$. Details of the matching algorithm can vary, but most studies, save for a few exceptions\autocite{polinski2022_durability, ioannou2022_effectiveness, gazit_incidence_2022}, have used 1:1 exact matching. We assume that successfully matched pairs from this form of matching are used to create a new dataset for analysis. For each individual within a matched pair, we define $V$ as the vaccination status of an individual on the day they were matched and $D$ as the day on which they were matched. Day $D$, which is also the day of vaccination for the vaccinated individual, is defined as the start of follow-up for both individuals, which prevents selection and immortal time bias. 
%The strategy of matching and assigning vaccination dates also ensures that the distribution of $X$ and $D$ are the same in the vaccinated and unvaccinated groups, thereby controlling for confounding by these factors. 
Then, assuming $X$ is a sufficient set of confounders, the matched data set can be analyzed as would be data from a randomized  trial, with two notable exceptions: (i) if either individual within a matched pair experiences the endpoint within $\tau$ days of $D$, then the pair is excluded from the analysis; (ii) if a matched control receives a vaccine after day $D$, then both individuals in the pair are considered right-censored on that day. Alternatively, one could choose to match on some, but not all, confounders to increase the probability of finding matches, and then appropriately adjust for additional known confounders in the analysis \autocite{sjolander2013_ignoring, mansournia2013_matcheda}.

The matching-based estimands are analogous to those used in randomized trials and, in the case of the first strategy, can be estimated in the same way. With slight abuse of notation, we use $T = Y - D$ to denote time in days from matching on day $D$ to first incidence of the study endpoint and let $R(\tau) = I(T > \tau)$. The matching-based marginal cumulative incidence has the same form as that for the randomized trial, $P(T\leq t_0 \mid V = v, R(\tau) = 1)$, but is implicitly defined with respect to the population formed by the matching procedure and included in the matching-based analysis. A matching vaccine effectiveness estimand can  be defined as $\text{VE}_{\text{M}}(t_0)= 1 - P(T\leq t_0 \mid V = 1, R(\tau) = 1) / P(T\leq t_0 \mid V = 0, R(\tau) = 1)$. 

As before, the marginal cumulative incidence can be written as a weighted combination of $D$- and $X$-specific cumulative incidences:
\begin{equation}
\begin{aligned} 
&P(T \leq t_0 \mid V = v, R^*(\tau) = 1) \\
&\hspace{2em} = \sum_{x,d} \bigg [ P(T \leq t_0 \mid V = v,  R^*(\tau) = 1, D = d, X = x) \\
&\hspace{7em} \times P(D = d \mid V = v, R^*(\tau) = 1, X = x) P(X = x \mid V = v, R^*(\tau) = 1) \bigg] \ .\label{eqn:cuminc_match}
\end{aligned}
\end{equation}
By the design of matching, the marginalizing distributions for vaccination dates ($ P(D = d \mid V = v, R(\tau) = 1, X = x))$ and covariates $(P(X = x \mid V = v,  R(\tau) = 1))$ are the same in the vaccinated and unvaccinated groups, thereby preventing confounding by these factors. These marginalizing distributions are defined in the subpopulation who naturally uptake vaccine and are included in the matched analysis; this may be an important distinction relative to the randomized trial setting if vaccine uptake varies by key characteristics (e.g., age).

%A distinction between $\text{VE}_{\text{RCT}}(t_0)$ and $\text{VE}_{\text{M}}(t_0)$ is that the implicit marginalizing covariate distribution used by the matching procedure is that of the subpopulation who naturally uptake vaccine, $P(X = x \mid V^* = 1, R^*(\tau) = 1)$, which may be an important distinction relative to the randomized trial setting if vaccine uptake varies by key characteristics (e.g., age).

\section{A new causal estimand}

The estimands above are presented as summaries of observed data distributions because their target causal estimands are often not explicitly defined. Here, we propose a general causal vaccine effectiveness estimand and describe how it can be identified and estimated. 
%This transparency allows one to scrutinized the assumptions needed for causal interpretation.
When describing the proposed estimand, we assume the single unit treatment value assumption (SUTVA) \autocite{rubin1980_randomization}, which stipulates there is only a single formulation of vaccine and that there is no interference between individuals. 

%the assumption is likely reasonable for studies in which enrolled individuals are a small subset of the population.
\subsection{Causal vaccine effectiveness}

Using target trial principles, we imagine a hypothetical experiment that starts on on calendar day $d_0$. We consider a \emph{joint intervention}, which assigns \emph{both} vaccine $v$  \emph{and} the date $d$ on which that vaccine is given. The joint intervention can also be conceptualized as a longitudinal intervention \autocite{hernan2020_causal} that consists of giving no vaccine from the study start $d_0$ until day $d$, and then if no endpoint has occurred, administering vaccine $v$ on day $d$, and no vaccine thereafter. The intervention also prevents censoring before the end of follow-up, although we leave this implicit in our notation. Under such an intervention, we would observe the counterfactual $Y(d,v)$, which describes the time from study start to endpoint under our intervention. An individual could experience the endpoint prior to their assigned vaccination date such that $Y(d,v) \leq d$.

A potential estimand of interest is $\psi_v(t_0; d,x) = P(Y(d,v) \leq  t_0 + d \mid Y(d,0) > d + \tau, Y(d,1) > d + \tau, X = x)$, which describes a conditional cumulative incidence in the principal stratum \autocite{frangakis2002_principal} of individuals who would remain endpoint free $\tau$ days beyond their assigned vaccination date irrespective of their assigned vaccine $v$. Conditioning on this principal stratum ensures that a comparison of $\psi_v(t_0; d,x)$ between active vaccine $v=1$ and no vaccine $v = 0$ describes a causal $d$- and $x$-specific vaccine effect.

As in (\ref{eqn:cuminc_rct}) and (\ref{eqn:cuminc_match}), we define a  summary measure by taking a weighted average of the $d$- and $x$-specific cumulative incidences, \begin{equation}
\bar{\psi}_v(t_0) = \sum_{x,d} \psi_v(t_0; d,x) g^*(d\mid x) p^*(x) \ ,   \label{eqn:psibar}
\end{equation}
where the marginalizing weights are user-specified functions $g^*$ and $p^*$, such that $\sum_{d} g^*(d \mid x) = 1$ for all $x$ and $\sum_x p^*(x) = 1$. The covariate-conditional weight given to each day $g^*(d\mid x)$ and the weight given to each covariate level $p^*(x)$ could be fixed or based on distributions in the observed data. A causal vaccine effectiveness estimand can be defined as $\text{VE}_{\text{C}}(t_0) = 1 - \bar{\psi}_1(t_0)/\bar{\psi}_0(t_0).$

While any appropriate weight functions $g^*$ and $p^*$ can be used in (\ref{eqn:psibar}), a choice of weights that can be easily estimated is \begin{equation}
\begin{aligned}
g^*(d \mid x) &= P(D^* = d \mid X = x, V^* = 1, Y - D^* > \tau) , \text{ and } \\ 
p^*(x) &= P(X = x \mid V^* = 1,  Y - D^* > \tau) \ ,
\end{aligned}
\label{eqn:observed_marginalization} 
\end{equation}
which provides close alignment of (\ref{eqn:psibar}) with the matching estimand (\ref{eqn:cuminc_match}). In particular, if all vaccinated individuals are matched, these weights are the same as those in the matching estimand. 

We note that, ignoring the additional conditioning on $x$, the proposed $\psi_v(t_0;d,x)$ is similar to the calendar-time-specific cumulative incidence used in the causal VE estimand of Demonte et al \autocite{demonte2024_assessing} and a ``trial-specific'' cumulative incidence in the sequential trials approach \autocite{hernan2008_observational}. An important distinction is that we have clearly defined the target population for $\psi_v(t_0;d,x)$, whereas the other estimands have conditioned on the population \textit{observed} to be eligible and at-risk, without further specifying the target population for inference. Our marginalized estimand in (\ref{eqn:psibar}) is similar to the estimand of a pooled sequential trials analysis but, importantly,  is explicit in how each  component is weighted. 
%Our estimand can be thought of as an explicit ``pooling'' of the well-defined hypothetical trials we described above.  

 \subsection{Identification}

We consider the same observed data structure as in Section \ref{sec:obs_studies_ve}, generalized to allow for right-censoring. We assume the data consist of $X$, $V^*$, $D^*$, and $(\tilde{Y}, \delta)$, where $\tilde{Y}$ is the minimum of the time to endpoint $Y$ and right-censoring time $C$, while $\delta = I(\tilde{Y} = Y)$ is an event indicator. We use $\tilde{C}_k = I(\tilde{Y} \le k, \delta = 0)$ to denote an indicator of right-censoring by day $k$ and $\tilde{Y}_k = I(\tilde{Y} \le k, \delta = 1)$ to denote an indicator of observing an endpoint by day $k$, where $k = K$ denotes the maximum follow-up time of interest. We define $V_k$ to be the type of vaccine received on day $k$ (0 = no vaccine; 1 = active vaccine). We assume the ordering  $(\tilde{C}_k, \tilde{Y}_k, V_k)$. We use bar notation to denote the history of a variable through a particular day, e.g., $\bar{V}_k = (V_1, V_2, \dots, V_k)$. We define $Y_k(d,v) = I(Y(d,v) \leq k)$ to denote the counterfactual outcome occurring by day $k$.  We also define $\nu_k^{d,v} = I(k =d, v = 1)$ to be the vaccine given on day $k$ that is consistent with the vaccine strategy ``give vaccine $v$ on day $d$ and no vaccine otherwise''; the full longitudinal vaccine strategy is represented by $\bar{\nu}_{\infty}^{d,v}$.

\noindent \textbf{Assumption 1: Exchangeability of vaccination and censoring given covariates and vaccination history}

\begin{align*}
%&(Y_{k+1}(d,v), \ldots,Y_K(d,v)) \perp (V_k, \tilde{C}_{k+1})  \mid X, \overline{V}_{k-1} = \bar{\nu}_{k-1}^{d,v}, \tilde{C}_k = 0, \tilde{Y}_k = 0 \ , \\
&(Y_{k+1}(d,v), \ldots,Y_K(d,v)) \perp V_k \mid X, \overline{V}_{k-1} = \bar{\nu}_{k-1}^{d,v}, \tilde{C}_k = 0, \tilde{Y}_k = 0 \ , \\
&(Y_{k+1}(d,v), \ldots,Y_K(d,v)) \perp \tilde{C}_{k+1} \mid X, \overline{V}_{k} = \bar{\nu}_{k}^{d,v}, \tilde{C}_k = 0, \tilde{Y}_k = 0 \ , \\
& \text{ for all $d$ and for } k = 1, \ldots,K-1.
\end{align*}

%Mats: suggestion:
% \begin{align*}
% &(Y_{k+1}(d,v,c=0), \ldots,Y_K(d,v,c=0)) \perp V_k \mid X, \overline{V}_{k-1} = \bar{\nu}_{k-1}^{d,v,c=0}, \tilde{C}_k = 0, \tilde{Y}_k = 0 \ , \\
% &(Y_{k+1}(d,v,c=0), \ldots,Y_K(d,v,c=0)) \perp \tilde{C}_{k+1} \mid X, \overline{V}_{k} = \bar{\nu}_{k}^{d,v,c=0}, \tilde{C}_k = 0, \tilde{Y}_k = 0 \ , \\
% & \text{ for all $d$ and for } k = 1, \ldots,K-1.
% \end{align*}

This assumption states that an at-risk individual's decision to get vaccinated and/or leave the study on any day is independent of their future outcomes given their covariates and prior vaccination history. This assumption requires that $X$ be a sufficiently rich collection of covariates to enable confounding control for both the vaccination and censoring process.

\noindent \textbf{Assumption 2: Positivity}
\begin{align*}
& \text{For any } x \text{ such that $p^*(x) > 0$ and for all combinations of $d$ and $v$}, \\
& 0 < P(V_k = \nu_k^{d,v}, \tilde{C}_{k+1} = 0 \mid X = x, \overline{V}_{k-1} = \bar{\nu}_{k-1}^{d,v}, \tilde{Y}_k = 0, \tilde{C}_k = 0) < 1 \ ,\\
& \text{ for all } k = 1,...,K-1.
\end{align*}
Assumption 2 states that on each day, there is a positive probability that an individual follows the vaccine intervention of interest within each covariate strata. 

%\noindent \textbf{Assumption 3: Consistency}
%\[\text{If }  \overline{V}_{k-1} = \bar{v}_{k-1}^{d,v} \text{ and } C_{k} = 0, \text{ then } \bar{Y}_k(d,v) = \bar{Y}_k \ .\]
%Assumption 3 states that if an individual's observed vaccination history through day $k -1$ is the same as the vaccination history that would be assigned by a vaccine intervention of interest through the same time, then the counterfactual endpoint history under that intervention equals the observed endpoint history through day $k$. Note that this also implies that the counterfactual event indicators $Y_k(d,v)$ for different vaccine interventions $(d,v)$ can be the same when the interventions involve the same vaccine assignment histories up to day $k$. 

\noindent \textbf{Assumption 3: No impact of intention to vaccinate and no vaccine effect until $\tau$ days  after vaccination}
\begin{align*}
& Y(d, 1) > d + \tau \iff Y(d, 0) > d + \tau \ , \text{ for all $d$}
\end{align*}
Assumption 3 states that the type of vaccine assigned on day $d$ does not impact an individual's risk for the endpoint until day $d + \tau$, 
%If $\tau = 0$, this is, in fact, (partially?) implied by the standard consistency assumption (Assumption 3). 
implying that individuals do not modify their risk behaviors in response to intention to be vaccinated in the future. The assumption also requires that the vaccine has no immunological impact on risk of the study endpoint until $\tau$ days after vaccination.

%\noindent \textbf{Assumption 4: No vaccine on day $d$ is equivalent to placebo/control vaccine on day $d$}
% \[ Y(d, 0) = Y(d,\emptyset) = Y(\underline{\emptyset})\]
%To compare observational effectiveness studies to randomized controlled trials, we must assume that receiving no vaccination is equivalent to receiving placebo/control vaccine in the context of a randomized trial. We also note that assigning no vaccine $v = \emptyset$ on day $d$ is equivalent to a treatment policy of ``never vaccinate," which we assume does not depend on vaccination day $d$. 
%This latter assumption is needed for identification since vaccination dates are only observed for those who receive the active vaccine. 
%Making inference about the causal effect of active vs. placebo vaccine would thus require assumptions about the equivalence between no vaccine and placebo vaccine. 

\noindent \textbf{Assumption 4: No vaccine on day $d$ is equivalent to never vaccinate}
\[ Y(d, 0) = Y(0) , \text{ for all $d$}\]
Assumption 4 emphasizes that we have defined the control intervention as ``no vaccine'' rather than placebo vaccine since placebo/control vaccines are not observed in observational data. In our conceptualization of the hypothetical intervention, giving no vaccine on day $d$ is equivalent to a treatment policy of ``never vaccinate''.  We define $Y(0)$ as the time from study start to endpoint if an individual is prevented from ever being vaccinated. 

As with many observational scenarios, the assumptions necessary for causal inference are strong. 
%Assumption 1 requires detailed covariate information on individuals and deconfounding of an inherently complex socio-behavioral process. Causal graphs can be useful for identifying sufficient sets of covariates for adjustment \autocite{greenland1999_causal, pearl2009_causality, spirtes2001_causation}; however, this assumption can never be fully verified in practice. Assumption 3 may also be challenging to justify in practice in some contexts. In some settings individuals may seek to minimize their risk behavior until they receive a vaccine (e.g., by delaying a vacation) and complete information on such risk behaviors is rarely available to an analyst.
In spite of the strength of these assumptions, articulating a sufficient set of assumptions for causal inference is useful for designing future studies. 
It also may shed light on the implicit assumptions required to interpret the results of matching analyses in a causal way. To the best of our knowledge, these assumptions have not been formally described; however, we believe they share considerable overlap with ours. For example, matching-based analyses typically treat the matched dataset as unconditionally randomized, which requires  assumptions of conditional exchangeability given the variables used for matching \autocite{stuart2010_matching}.
%From a practical standpoint, 
Matching also enforces positivity because individuals in strata with positivity violations will not be able to be matched and will therefore be excluded from the analysis \autocite{hernan2020_causal}. Assumption 3 motivates the matching practice of including only matched pairs in which both individuals are at risk $\tau$ days after $D$. Overall, we conjecture that the assumptions required to identify our proposed estimand are similar to those required to imbue matching with a causal interpretation.

Our identification result relies on identification of two conditional hazard functions: 1) the conditional hazard for the time to study endpoint from study start $d_0$ among individuals not yet vaccinated,
\[\lambda_0(t; x) = P(\tilde{Y} = t  \mid  \tilde{Y} > t-1,  \bar{V}_{t-1} = \bm{0}_{t-1}, \tilde{C}_{t} = 0, X = x) \  
\ , \] 
and 2) the conditional hazard for the time to study endpoint from vaccination time $D^*$ among vaccinated individuals,  
\begin{align*}
\lambda_1(t; d, x) %&= P[Y - d = t  \mid Y - d > t-1,  D_{obs} = d, C_{d + t} = 0, X] \\ 
&= P(\tilde{T} = t  \mid \tilde{T} > t-1,  D^* = d, \tilde{C}_{d + t} = 0,  X = x) \ ,
\end{align*} 
where $\bm{0}_{t-1}$ is a vector of zeroes of length $t-1$ and $\tilde{T} = \tilde{Y} - D^*$. 

\begin{theorem}
 Under Assumptions 1-4 and SUTVA,
\begin{equation}
 \psi_0(t_0; d, x) = 1 - \prod_{s=d + \tau + 1}^{d + t_0} \{1 -  {\lambda_0(s; x)}\} \ , 
\label{eqn:identification_result_psi0}
\end{equation}
and 
\begin{equation}
 \psi_1(t_0; d,x) =1 - \prod_{s = \tau + 1}^{t_0}\{1 -  {\lambda_1(s; d, x)}  \} \ .
\label{eqn:identification_result_psi1}
\end{equation}
\end{theorem}
%This proposition states that the counterfactual cumulative incidences are identified by functions of the two observed hazards. 
See eAppendix 1 for a proof. 

\subsection{Estimation}

The identification formulas suggest that estimators for $\text{VE}_\text{C}(t_0)$ can be computed based on estimators of conditional hazards. While many such methods are available, we explicitly describe an estimator based on Cox proportional hazards models.

\begin{enumerate}
\item To estimate $\lambda_0(t; x)$, fit a Cox model where days since $d_0$ is the time scale and $X$ is included in the model formula. All individuals are included and individuals' data are considered right-censored at the minimum of their observed day of vaccination, their right-censoring date, and the study end date. An estimate of $\lambda_0(t; x)$ is given by combining the regression coefficients with the Nelson-Aalen baseline hazard estimate. 

\item To estimate $\lambda_1(t; d, x)$, fit a Cox regression model including only individuals who were vaccinated during the study time period and who did not experience the study endpoint within $\tau$ days of vaccination. The time scale for this model is days since vaccination and individuals are considered right-censored at the minimum of their right-censoring date and the study end date. The model formula should  adjust for $X$ and $D^*$. The latter could be achieved by specifying a  flexible form of $D^*$ (e.g., penalized cubic splines). 

\item Compute plug-in estimators of $\bar{\psi}_0(t_0)$ and $\bar{\psi}_1(t_0)$ based on the selected marginalizing weights $g^*$ and $p^*$. If one selects $g^*$ and $p^*$ as suggested in (\ref{eqn:observed_marginalization}), then denoting the set of vaccinated individuals who remained at-risk $\tau$ days after vaccination by $\mathcal{V}(\tau) =\{i: V_i^*\times I(Y_i - D^*_i>\tau)\}$ and the number of individuals in the set by $|\cdot|$, estimates can be computed as \begin{align*}
        \hat{\bar{\psi}}_0(t_0) &= \frac{1}{|\mathcal{V}(\tau)|} \sum_{i \in  \mathcal{V}(\tau)}  \left[1 - \prod_{t = D^*_i + \tau + 1}^{D^*_i + t_0}\{1 - \hat{\lambda}_0(t; X_i)\}\right] \\ 
        \hat{\bar{\psi}}_1(t_0) &= \frac{1}{|\mathcal{V}(\tau)|} \sum_{i \in \mathcal{V}(\tau)} \left[1 - \prod_{t = \tau + 1}^{t_0} \{1 - \hat{\lambda}_1(t; D^*_i, X_i)\}\right]  \ , 
\end{align*}
The plug-in estimator of vaccine effectiveness is $\widehat{\text{VE}}_{\text{C}}(t_0) = 1 - \hat{\bar{\psi}}_1(t_0)/\hat{\bar{\psi}}_0(t_0)$. 
\item If incidence and/or vaccine efficacy curves are of interest, repeat step 3 for all times $t_0$ of interest.
\end{enumerate}

We recommend using the nonparametric bootstrap for generating pointwise and simultaneous confidence intervals for the vaccine effectiveness curve (see eAppendix 2 for details). For cumulative incidence, we recommend creating confidence intervals on the logit scale before back-transforming; for $\text{VE}$, we recommend creating confidence intervals on the log risk-ratio scale, $\mbox{log}(1 - \text{VE})$, before back-transforming. Alternatively, percentile bootstrap methods can be applied.

\section{Simulation}
We evaluated the performance of the proposed estimator relative to matching via simulation (see eTable1 for details). The simulation design was chosen to resemble the dataset from the study described below examining vaccine effectiveness for preventing SARS-CoV2 infection in a large school district in the United States. We simulated a four-dimensional covariate $X$, binary vaccine status, timing of vaccination, timing of the infection endpoint of interest, and right-censoring time. We allowed the effect of vaccination and the baseline risk of infection to change over calendar time.

We compared our estimator to a matching estimator in terms of bias, mean squared error (MSE), coverage of a nominal 95\% bootstrap-based confidence interval, confidence interval width, and relative efficiency of the point estimates based on the ratio of MSEs. Our estimator relied on using two Cox proportional hazards regression models, as previously described. The hazard model for the vaccinated included a natural penalized cubic spline for $D^*$ with default knot values. Wald-style bootstrap confidence intervals were computed using 1000 bootstrap resamples.

For the matching estimator, we conducted rolling-cohort matching with 1:1 exact matching on $X$. We estimated the matching-based marginal cumulative incidences and VE using Kaplan-Meier. We considered estimating these quantities using Cox regression models, which yielded similar results (eFigure 1). We constructed Wald-style bootstrap confidence intervals by resampling matched pairs, keeping the initial matched dataset fixed. We also considered bootstrapping the original observed data and then rematching, which yielded similar results (eFigure 1). 

For estimation of VE, both methods showed little bias and coverage close to 95\% across all sample sizes (Table \ref{tab:sim_results}). The exception is the matching-based estimator at a sample size of $N = 500$, which showed higher bias and lower coverage due to some extreme estimates. The proposed estimator outperformed the matching-based estimator in terms of precision with significantly narrower confidence intervals and relative efficiency gains of 40\% to 86\%. The precision gains for the proposed estimator were greatest at the smaller sample sizes but were substantial at the larger sample sizes as well. These results were also seen for the cumulative incidence terms used to compute VE (eTable 2).

\begin{table}
\centering
\caption{Simulation study results for estimation of VE based on 1000 simulations}
\centering
\begin{threeparttable}
\begin{tabular}[t]{rlccccc}
\toprule
\textbf{N} & \textbf{Method} & \textbf{Bias} & \textbf{MSE} & \textbf{Coverage} & \textbf{Width} & \textbf{Rel.Eff.}\\
\midrule
 & matching & -0.157 & 0.654 & 0.903 & 2.736 & 1.000\\

\multirow[t]{-2}{*}{\raggedleft\arraybackslash 500} & proposed & -0.022 & 0.092 & 0.967 & 2.176 & 0.141\\
\cmidrule{1-7}
 & matching & -0.043 & 0.111 & 0.974 & 1.941 & 1.000\\

\multirow[t]{-2}{*}{\raggedleft\arraybackslash 1000} & proposed & -0.020 & 0.039 & 0.972 & 1.317 & 0.355\\
\cmidrule{1-7}
 & matching & -0.014 & 0.036 & 0.961 & 1.181 & 1.000\\

\multirow[t]{-2}{*}{\raggedleft\arraybackslash 2000} & proposed & -0.010 & 0.021 & 0.952 & 0.878 & 0.584\\
\cmidrule{1-7}
 & matching & -0.005 & 0.012 & 0.950 & 0.688 & 1.000\\

\multirow[t]{-2}{*}{\raggedleft\arraybackslash 5000} & proposed & -0.004 & 0.007 & 0.948 & 0.540 & 0.606\\
\bottomrule
\end{tabular}
\begin{tablenotes}
\item MSE = mean squared error; Coverage = coverage of nominal 95\% Wald-style bootstrap confidence intervals; Width = average confidence interval width; Rel.Eff. = ratio of the MSE of proposed estimator vs. matching estimator. Both Width and Rel.Eff. are on the log(1-VE) scale. The true value of VE in this scenario was 0.38.
\end{tablenotes}
\end{threeparttable}
\label{tab:sim_results}
\end{table}

\section{Application}
 We illustrate the proposed method using a study  on the effectiveness of the Pfizer-BioNTech COVID-19 vaccine among children aged 5-11 years old during the 2021-2022 school year \autocite{harton2025_estimating}. The data come from a large urban school district in the United States that implemented an opt-in testing program in which students could receive a COVID-19 antigen test at least once a week. Our outcome of interest is test-confirmed SARS-CoV-2 infection $\tau$=14 days after receiving the first dose of the vaccine. The study period was  October 29, 2021, the date children aged 5-11 years old became eligible for the Pfizer-BioNTech COVID-19 vaccine,  to May 26, 2022, the end of the school year.

The analytic cohort consisted of 9209 students from 55 schools in ten school clusters. 41\% of students were vaccinated prior to first infection, with the majority of vaccinations occurring shortly after becoming eligible. The median time to receiving vaccination from study start was 15 days (IQR: 10 to 35, Range: 4 to 207). 
%The vaccine-uptake period coincided with the emergence of the Omicron variant and its rapid replacement of the previously circulating Delta variant of SARS-CoV2 \autocite{team2021_sarscov2}. Thus, we expect from previous studies that vaccine effectiveness may change over the observed dates of vaccination \autocite{khan2022_estimated, sabu2022_effectiveness, buchan2022_estimated}. 
Students were right-censored if they missed four consecutive tests during the school year, with their date of right-censoring assigned to be the date of their last recorded test. A small percentage (2.6\%) of students were additionally right-censored due to disenrollment from school. The median follow-up time was 206 days (IQR: 187 to 207) with a maximum possible follow-up time of 209 days.  

We applied the proposed and matching-based methods to evaluate cumulative incidence and vaccine effectiveness at $t_0 = 15,16,...,180$ days since vaccination.  For both methods, we adjusted for covariates of age, sex, race, and school cluster. After matching on the selected variables, the matched cohort was 65\% of the original sample size. 

\begin{figure}
    \centering
\includegraphics[width=0.9\linewidth]{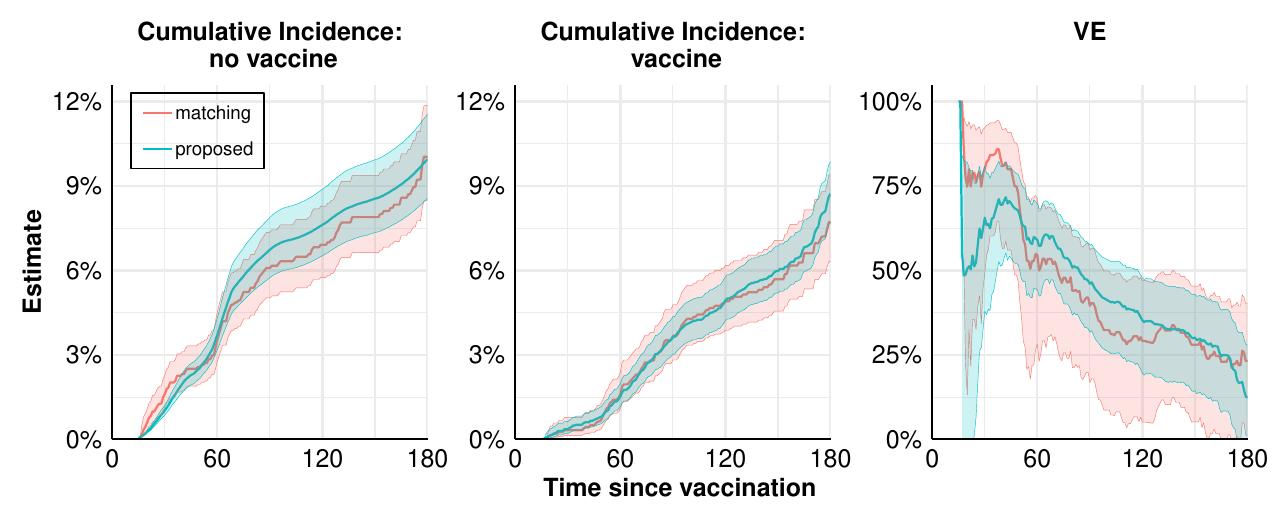}
    \caption{Cumulative incidence of SARS-CoV-2 infection in children 5-11 years old and effectiveness of the Pfizer-BioNTech COVID-19 vaccine over time. Shaded areas represent 95\% pointwise Wald-style confidence intervals based on 1000 bootstrap resamples. }
    \label{fig:ve}
\end{figure}

\begin{figure}
    \centering
    \includegraphics[width=0.8\linewidth]{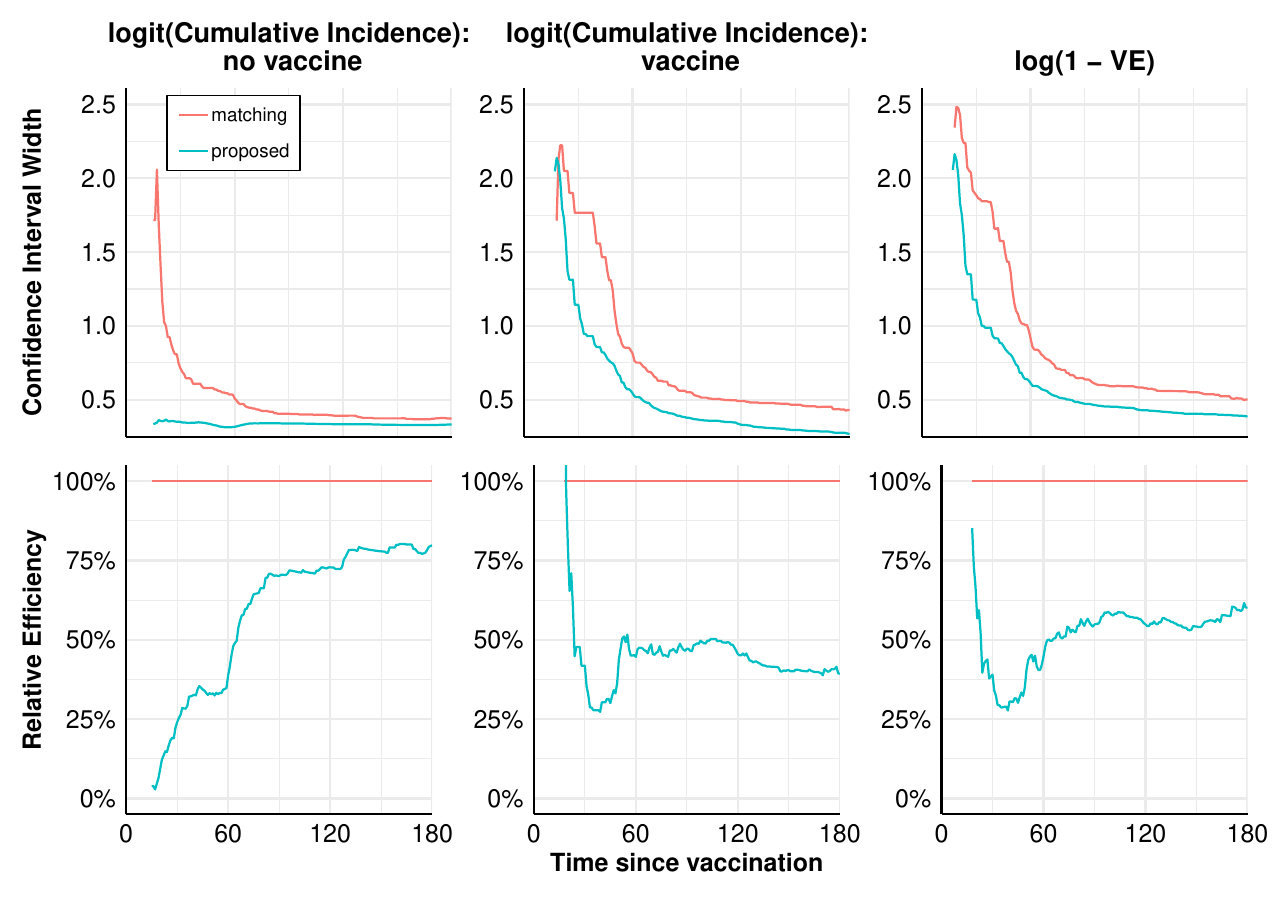}
    \caption{Confidence interval width and relative efficiency of matching and proposed estimators in illustrative study on the indicated scales. Relative efficiency is defined as the ratio of bootstrap variances.}
    \label{fig:precision_gain}
\end{figure}

The proposed and matching methods provided similar point estimates for cumulative incidence in the vaccinated and unvaccinated groups and for vaccine effectiveness (Figure \ref{fig:ve}). Both methods showed waning vaccine effectiveness  over time. For example, estimates of VE imply moderate protection at $t_0 = 60$ (Matching = 0.55 (95\% CI: 0.28 to 0.71); Proposed = 0.59 (95\% CI: 0.44 to 0.70)) but little to no protection at $t_0 = 180$ days after vaccination (Matching = 0.23 (95\% CI: 0.01 to 0.40); Proposed = 0.12 (95\% CI: -0.07 to 0.28). The point-wise confidence intervals for the proposed estimator were considerably narrower than those for matching at almost all timepoints. (Figure \ref{fig:precision_gain}). The same result was seen for simultaneous confidence intervals (eFigure 3), where narrower confidence intervals for VE based on the proposed method tended to exclude the null at more timepoints than those for matching. %We found that the proposed method yielded evidence of cumulative vaccine effectiveness between 30 and 165 days after vaccination (as evidenced by the simultaneous confidence band excluding zero effectiveness), while the simultaneous confidence band based on matching indicates evidence of effectiveness only between 30 and 95 days after vaccination.

In addition to confidence interval width, we studied the relative efficiency of the estimators by comparing the ratio of their estimated variances (Figure \ref{fig:precision_gain}). The proposed method yielded efficiency gains of 40-61\% over matching for the estimation of VE, with larger efficiency gains at earlier follow-up times. 

We note that our results based on matching where matches were randomly selected from all eligible matches led to considerable dependence of the results on the initial random seed used (eFigure 2). 

% Discussion point? Previous pooled COVID-19 VE estimates among children aged 5-11 years against any SAR-CoV-2 infection during the Omicron period were 46.3\% (Li) and 41.6 (Piechotta)\%.

\section{Discussion}
%We formalized a novel causal estimand for observational studies of vaccine effectiveness using a tar get trial emulation framework, and in doing so highlight limitations to the estimand for the current matching approach which does not have a formal causal interpretation.%The proposed approach shows promise as an alternative to matching-based analyses of vaccine effectiveness both because of the implications for causal inference and the statistical efficiency gains of the estimation procedure.
%Even though our simulation and data settings considered a relatively common endpoint (approximately 10% cumu lative incidence in placebo group), the proposed method produced substantial efficiency gains. We expect our approach to be even more useful for evaluating effectiveness of vaccines against rare events, such as death and hospitalization.

%In doing so, we highlight the lack of clarity about the target causal estimand in current matching approaches.  The proposed approach is advantageous in that it provides a clear framework for causal interpretation and enables greater statistical efficiency in the estimation procedure.

We formalized a novel causal estimand of vaccine effectiveness that can be used as an alternative to matching in observational studies using target trial emulation. The proposed approach provides a clear framework for causal interpretation and highlights some lack of clarity about the target causal estimand in matching-based and sequential trial approaches. By first articulating our causal estimand, we were also able to develop a straight-forward and efficient estimation approach \autocite{dang2023_start}.  Our simple estimation approach only requires fitting two hazard-based regressions -- methods that most analysts are accustomed to working with.  More importantly, we observed substantial efficiency gains in real and simulated settings with a relatively common endpoint (approximately 10\% cumulative incidence in placebo group). We expect our approach will be even more useful for evaluating effectiveness of vaccines against rare but important endpoints, such as death and hospitalization. 

The proposed method may also obviate some challenges of matching including making bias-variance tradeoffs in deciding what variables to match on and checking covariate balance after each matching attempt\autocite{ho2007_matching, iacus2012_causal}. In addition, similar to others, we found that matching can depend on the random seed set prior to matching \autocite{andrillon2020_performance, monteiro2023_impact}. This suggests that matching should include running and averaging over several iterations to obtain stable inference; combined with bootstrapping, this may yield a computationally intensive analysis. 
%Matching may also present challenges to an analyst when it is infeasible to match on the full set of identified confounders $X$, as doing so would result in an unacceptable amount of unmatched (and thus discarded) data. An analyst must then tune a challenging bias-variance trade-off by selecting which variables to match on which to adjust for in the analysis phase. Regression modeling simplifies the tuning of the bias-variance trade-off and potentially enables the use of straightforward approaches for estimator selection, such as cross-validation, although more complex estimators would be needed to ensure valid inference (e.g., \citet{benkeser2019estimating}).

While our approach offers important improvements over matching, it shares some of its limitations. First, we assume that baseline covariates measured at study start are sufficient to explain vaccination timing. In practice, vaccination timing may be influenced by time-varying factors such as community transmission levels. When such information is available, it should be possible to extend our methods to account for this using more complex formulations of hazard regression models. Although matching appears to allow adjustment for time-varying confounders of vaccination by defining baseline covariates at the time of matching, standard analyses of matched data will not fully account for time-varying factors of artificial censoring due to vaccination. Another shared limitation is our assumption of no interference, which may not hold in infectious disease contexts \autocite{hudgens2008_causal, halloran2016_dependent}. However, we expect that allowing partial interference is feasible within our estimation framework \autocite{sobel2006_what, tchetgen2012_causal}. 
%For example, in our data analysis, detailed classroom information was not made available due to privacy concerns. However, if such information were available, then a similar analysis could be undertaken using classroom as the unit of observation and under a partial interference structure where only the fraction of an individual's classmates who are vaccinated influences their counterfactual outcome \autocite{hong2006_evaluating}. In this approach, it may be possible to construct a time-varying, classroom-level covariate describing the fraction of the class that has been vaccinated and to include this variable in the modeling. 

To facilitate practical application, we have developed the \texttt{nomatchVE R} package 
implementing the estimators used in this paper, available at \url{https://github.com/ewu16/nomatchVE}.

\printbibliography

@article{mehrotra2021clinical,
  title={Clinical endpoints for evaluating efficacy in COVID-19 vaccine trials},
  author={Mehrotra, Devan V and Janes, Holly E and Fleming, Thomas R and Annunziato, Paula W and Neuzil, Kathleen M and Carpp, Lindsay N and Benkeser, David and Brown, Elizabeth R and Carone, Marco and Cho, Iksung and others},
  journal={Annals of Internal Medicine},
  volume={174},
  number={2},
  pages={221--228},
  year={2021},
  publisher={American College of Physicians}
}

@article{cohen-stavichandraj.2022_bnt162b2,
    author = {Chandra J. Cohen-Stavi  and Ori Magen  and Noam Barda  and Shlomit Yaron  and Alon Peretz  and Doron Netzer  and Carlo Giaquinto  and Ali Judd  and Leonard Leibovici  and Miguel A. Hernán  and Marc Lipsitch  and Ben Y. Reis  and Ran D. Balicer  and Noa Dagan },
    title = {BNT162b2 Vaccine Effectiveness against Omicron in Children 5 to 11 Years of Age},
    journal = {New England Journal of Medicine},
    volume = {387},
    number = {3},
    pages = {227-236},
    year = {2022},
    doi = {10.1056/NEJMoa2205011},
    URL = {https://www.nejm.org/doi/full/10.1056/NEJMoa2205011},
    eprint = {https://www.nejm.org/doi/pdf/10.1056/NEJMoa2205011}
}

@article{dagannoa2021_bnt162b2,
    author = {Noa Dagan  and Noam Barda and Eldad Kepten  and Oren Miron  and Shay Perchik  and Mark A. Katz  and Miguel A. Hernán  and Marc Lipsitch  and Ben Reis  and Ran D. Balicer },
    title = {BNT162b2 mRNA Covid-19 Vaccine in a Nationwide Mass Vaccination Setting},
    journal = {New England Journal of Medicine},
    volume = {384},
    number = {15},
    pages = {1412-1423},
    year = {2021},
    doi = {10.1056/NEJMoa2101765},
    URL = {https://www.nejm.org/doi/full/10.1056/NEJMoa2101765},
    eprint = {https://www.nejm.org/doi/pdf/10.1056/NEJMoa2101765}
}

@article{dang2023_start,
  title={Start with the target trial protocol, then follow the roadmap for causal inference},
  author={Dang, Lauren E and Balzer, Laura B},
  journal={Epidemiology},
  volume={34},
  number={5},
  pages={619--623},
  year={2023},
  publisher={LWW},
  doi = {10.1097/EDE.0000000000001637}
}

@article{dean2018_design,
 title={Design of vaccine trials during outbreaks with and without a delayed vaccination comparator},
  author={Dean, Natalie E and Halloran, M Elizabeth and Longini, Ira M},
  journal={The Annals of Applied Statistics},
  doi = {10.1214/17-AOAS1095},
  volume={12},
  number={1},
  pages={330},
  year={2018}
}

@article{dean2019_design,
    author = {Natalie E. Dean  and Pierre-Stéphane Gsell  and Ron Brookmeyer  and Victor De Gruttola  and Christl A. Donnelly  and M. Elizabeth Halloran  and Momodou Jasseh  and Martha Nason  and Ximena Riveros  and Conall H. Watson  and Ana Maria Henao-Restrepo  and Ira M. Longini },
    title = {Design of vaccine efficacy trials during public health emergencies},
    journal = {Science Translational Medicine},
    volume = {11},
    number = {499},
    pages = {eaat0360},
    year = {2019},
    doi = {10.1126/scitranslmed.aat0360},
    URL = {https://www.science.org/doi/abs/10.1126/scitranslmed.aat0360},
    eprint = {https://www.science.org/doi/pdf/10.1126/scitranslmed.aat0360}
}

@article{greenland1990_matching,
  title = {Matching and efficiency in cohort studies},
  author = {Greenland, Sander and Morgenstern, Hal},
  year = {1990},
  month = jan,
  journal = {American Journal of Epidemiology},
  volume = {131},
  number = {1},
  pages = {151--159},
  issn = {1476-6256, 0002-9262},
  doi = {10.1093/oxfordjournals.aje.a115469},
  urldate = {2024-05-06},
  langid = {english}
}

@article{hernan2016_specifying,
  title = {Specifying a Target Trial Prevents Immortal Time Bias and Other Self-Inflicted Injuries in Observational Analyses},
  author = {Hern{\'a}n, Miguel A. and Sauer, Brian C. and {Hern{\'a}ndez-D{\'i}az}, Sonia and Platt, Robert and Shrier, Ian},
  year = {2016},
  month = nov,
  journal = {Journal of Clinical Epidemiology},
  volume = {79},
  pages = {70--75},
  publisher = {Elsevier},
  issn = {0895-4356, 1878-5921},
  doi = {10.1016/j.jclinepi.2016.04.014},
  urldate = {2024-01-17},
  langid = {english},
  pmid = {27237061}
}

@article{hernan2016_using,
 author = {Hernán, Miguel A. and Robins, James M.},
    title = {Using Big Data to Emulate a Target Trial When a Randomized Trial Is Not Available},
    journal = {American Journal of Epidemiology},
    volume = {183},
    number = {8},
    pages = {758-764},
    year = {2016},
    month = {03},
    issn = {0002-9262},
    doi = {10.1093/aje/kwv254},
    url = {https://doi.org/10.1093/aje/kwv254},
    eprint = {https://academic.oup.com/aje/article-pdf/183/8/758/6652570/kwv254.pdf},
}

@article{hofner2024_vaccine,
    author = {Benjamin Hofner and Elina Asikanius and Wolfgang Jacquet and Theodor Framke and Katrien Oude Rengerink and Lukas Aguirre Dávila and Maria Grünewald and Florian Klinglmüller and Martin Posch and Finbarr P. Leacy and Thomas Lang and Armin Koch and Jörg Zinserling and Kit Roesextra and},
    title = {Vaccine Development during a Pandemic: General Lessons for Clinical Trial Design},
    journal = {Statistics in Biopharmaceutical Research},
    volume = {16},
    number = {2},
    pages = {158--170},
    year = {2024},
    publisher = {ASA Website},
    doi = {10.1080/19466315.2023.2211538},
    URL ={https://doi.org/10.1080/19466315.2023.2211538},
    eprint = { https://doi.org/10.1080/19466315.2023.2211538}
}

@article{hudgens2004_endpoints,
  title = {Endpoints in Vaccine Trials},
  author = {Hudgens, Michael G and Gilbert, Peter B and Self, Steven G},
  year = {2004},
  month = apr,
  journal = {Statistical Methods in Medical Research},
  volume = {13},
  number = {2},
  pages = {89--114},
  publisher = {SAGE Publications Ltd STM},
  issn = {0962-2802},
  doi = {10.1191/0962280204sm356ra},
  urldate = {2024-09-19}
}

@inbook{rose2011_open,
  title = {The Open Problem},
  booktitle = {Targeted {{Learning}}: {{Causal Inference}} for {{Observational}} and {{Experimental Data}}},
  author = {Rose, Sherri and {van der Laan}, Mark J.},
  editor = {{van der Laan}, Mark J. and Rose, Sherri},
  year = {2011},
  pages = {3--20},
  publisher = {Springer},
  address = {New York, NY},
  isbn = {978-1-4419-9782-1},
  langid = {english}
}

@article{stuart2010_matching,
  title={Matching methods for causal inference: A review and a look forward},
  author = {Stuart, Elizabeth A.},
  year = {2010},
  month = feb,
  journal={Stat Sci},
  shortjournal={Statistical science: a review journal of the Institute of Mathematifcal Statistics},
  volume = {25},
  number = {1},
  pages = {1--21},
  publisher = {Institute of Mathematical Statistics},
  issn = {0883-4237, 2168-8745},
  doi = {10.1214/09-STS313},
  urldate = {2024-06-26}
}

@article{westreich2010_invited,
   author = {Westreich, Daniel and Cole, Stephen R.},
    title = {Invited Commentary: Positivity in Practice},
    journal = {American Journal of Epidemiology},
    volume = {171},
    number = {6},
    pages = {674-677},
    year = {2010},
    month = {02},
    issn = {0002-9262},
    doi = {10.1093/aje/kwp436},
    url = {https://doi.org/10.1093/aje/kwp436},
    eprint = {https://academic.oup.com/aje/article-pdf/171/6/674/263877/kwp436.pdf},
}

@article{frangakis2002_principal,
    title={Principal stratification in causal inference},
    author = {Frangakis, Constantine E. and Rubin, Donald B.},
    journal = {Biometrics},
    number = {1},
    pages = {21--29},
    month = mar,
    year = {2002},
    volume = {58},
    issn = {0006-341X},
    url = {https://doi.org/10.1111/j.0006-341X.2002.00021.x},
    doi = {10.1111/j.0006-341X.2002.00021.x},
    urldate = {2024-04-11},
    publisher={Oxford University Press}
}

@article{suissa2023_prevalent,
    title={The prevalent new-user design for studies with no active comparator: the example of statins and cancer},
     author={Suissa, Samy and Dell’Aniello, Sophie and Renoux, Christel},
     journal={Epidemiology},
     volume={34},
     number={5},
     pages={681--689},
     year={2023},
     publisher={LWW},
    issn = {1044-3983},
    url = {https://journals.lww.com/epidem/abstract/2023/09000/the_prevalent_new_user_design_for_studies_with_no.11.aspx},
    doi = {10.1097/EDE.0000000000001628},
    language = {en-US},
    urldate = {2024-09-17}
}

@article{rubin1980_randomization,
    title={Randomization analysis of experimental data: The {Fisher} randomization test comment},
    author={Rubin, Donald B},
    journal={Journal of the American Statistical Association},
    volume={75},
    number={371},
    pages={591--593},
    year={1980},
    publisher={JSTOR},
    issn = {0162-1459},
    url = {https://www.jstor.org/stable/2287653},
    doi = {10.2307/2287653},
    urldate = {2025-02-24},
}

@article{hudgens2008_causal,
    title = {Toward Causal Inference With Interference},
    volume = {103},
    issn = {0162-1459},
    url = {https://doi.org/10.1198/016214508000000292},
    doi = {10.1198/016214508000000292},
    number = {482},
    urldate = {2025-02-24},
    journal = {Journal of the American Statistical Association},
    author = {Hudgens, Michael G and Halloran, M. Elizabeth},
    month = jun,
    year = {2008},
    pmid = {19081744},
    pages = {832--842},
}

@incollection{ruppert2003_inference,
    address = {Cambridge},
    series = {Cambridge {Series} in {Statistical} and {Probabilistic} {Mathematics}},
    title = {Inference},
    isbn = {978-0-521-78516-7},
    url = {https://www.cambridge.org/core/books/semiparametric-regression/inference/DE629458C17A4E772B59E8C6203381A5},
    urldate = {2025-02-25},
    booktitle = {Semiparametric {Regression}},
    publisher = {Cambridge University Press},
    author = {Ruppert, David and Wand, M. P. and Carroll, R. J.},
    year = {2003},
    doi = {10.1017/CBO9780511755453.008},
    pages = {133--160},
}

@article{halloran2016_dependent,
    title = {Dependent {Happenings}: a {Recent} {Methodological} {Review}},
    volume = {3},
    issn = {2196-2995},
    shorttitle = {Dependent {Happenings}},
    url = {http://link.springer.com/10.1007/s40471-016-0086-4},
    doi = {10.1007/s40471-016-0086-4},
    language = {en},
    number = {4},
    urldate = {2025-03-03},
    journal = {Current Epidemiology Reports},
    author = {Halloran, M. Elizabeth and Hudgens, Michael G.},
    month = dec,
    year = {2016},
    pages = {297--305},
}

@article{sobel2006_what,
    title = {What {Do} {Randomized} {Studies} of {Housing} {Mobility} {Demonstrate}?: {Causal} {Inference} in the {Face} of {Interference}},
    volume = {101},
    issn = {0162-1459},
    shorttitle = {What {Do} {Randomized} {Studies} of {Housing} {Mobility} {Demonstrate}?},
    url = {https://doi.org/10.1198/016214506000000636},
    doi = {10.1198/016214506000000636},
    number = {476},
    urldate = {2025-03-03},
    journal = {Journal of the American Statistical Association},
    author = {Sobel, Michael E},
    month = dec,
    year = {2006},
    publisher = {ASA Website}, 
    pages = {1398--1407},
}

@article{tchetgen2012_causal,
    title = {On causal inference in the presence of interference},
    volume = {21},
    issn = {0962-2802, 1477-0334},
    url = {https://journals.sagepub.com/doi/10.1177/0962280210386779},
    doi = {10.1177/0962280210386779},
    language = {en},
    number = {1},
    urldate = {2025-03-03},
    journal = {Statistical Methods in Medical Research},
    author = {{Tchetgen Tchetgen}, Eric J and VanderWeele, Tyler J},
    month = feb,
    year = {2012},
    pages = {55--75},
}

@article{pearce2023_are,
    title = {Are {Target} {Trial} {Emulations} the {Gold} {Standard} for {Observational} {Studies}?},
    volume = {34},
    issn = {1044-3983},
    url = {https://journals.lww.com/epidem/fulltext/2023/09000/are_target_trial_emulations_the_gold_standard_for.2.aspx},
    doi = {10.1097/EDE.0000000000001636},
    language = {en-US},
    number = {5},
    urldate = {2024-04-26},
    journal = {Epidemiology},
    author = {Pearce, Neil and Vandenbroucke, Jan P.},
    month = sep,
    year = {2023},
    pages = {614},
}

@article{barda2021_effectiveness,
    title = {Effectiveness of a third dose of the {BNT162b2} {mRNA} {COVID}-19 vaccine for preventing severe outcomes in {Israel}: an observational study},
    volume = {398},
    issn = {0140-6736, 1474-547X},
    shorttitle = {Effectiveness of a third dose of the {BNT162b2} {mRNA} {COVID}-19 vaccine for preventing severe outcomes in {Israel}},
    url = {https://www.thelancet.com/journals/lancet/article/PIIS0140-6736(21)02249-2/fulltext},
    doi = {10.1016/S0140-6736(21)02249-2},
    language = {English},
    number = {10316},
    urldate = {2024-04-26},
    journal = {The Lancet},
    author = {Barda, Noam and Dagan, Noa and Cohen, Cyrille and Hernán, Miguel A. and Lipsitch, Marc and Kohane, Isaac S. and Reis, Ben Y. and Balicer, Ran D.},
    month = dec,
    year = {2021},
    pmid = {34756184},
    publisher = {Elsevier},
    pages = {2093--2100},
}

@article{monge2023_effectiveness,
    title = {Effectiveness of a {Second} {Dose} of an {mRNA} {Vaccine} {Against} {Severe} {Acute} {Respiratory} {Syndrome} {Coronavirus} 2 ({SARS}-{CoV}-2) {Omicron} {Infection} in {Individuals} {Previously} {Infected} by {Other} {Variants}},
    volume = {76},
    issn = {1058-4838},
    url = {https://doi.org/10.1093/cid/ciac429},
    doi = {10.1093/cid/ciac429},
    number = {3},
    urldate = {2024-04-26},
    journal = {Clinical Infectious Diseases},
    author = {Monge, Susana and Rojas-Benedicto, Ayelén and Olmedo, Carmen and Martín-Merino, Elisa and Mazagatos, Clara and Limia, Aurora and Sierra, María José and Larrauri, Amparo and Hernán, Miguel A and {IBERCovid}},
    month = feb,
    year = {2023},
    pages = {e367--e374},
}

@article{lipsitch2016_observational,
    title = {Observational studies and the difficult quest for causality: lessons from vaccine effectiveness and impact studies},
    volume = {45},
    issn = {0300-5771},
    shorttitle = {Observational studies and the difficult quest for causality},
    url = {https://doi.org/10.1093/ije/dyw124},
    doi = {10.1093/ije/dyw124},
    number = {6},
    urldate = {2024-04-28},
    journal = {International Journal of Epidemiology},
    author = {Lipsitch, Marc and Jha, Ayan and Simonsen, Lone},
    month = dec,
    year = {2016},
    pages = {2060--2074},
}

@article{shiba2021_using,
    title = {Using {Propensity} {Scores} for {Causal} {Inference}: {Pitfalls} and {Tips}},
    volume = {31},
    issn = {0917-5040, 1349-9092},
    shorttitle = {Using {Propensity} {Scores} for {Causal} {Inference}},
    url = {https://www.jstage.jst.go.jp/article/jea/31/8/31_JE20210145/_article},
    doi = {10.2188/jea.JE20210145},
    language = {en},
    number = {8},
    urldate = {2024-06-03},
    journal = {Journal of Epidemiology},
    author = {Shiba, Koichiro and Kawahara, Takuya},
    month = aug,
    year = {2021},
    pages = {457--463},
}

@article{didelez2024_re,
    title = {Re: {Are} {Target} {Trial} {Emulations} the {Gold} {Standard} for {Observational} {Studies}?},
    volume = {35},
    issn = {1044-3983},
    shorttitle = {Re},
    url = {https://journals.lww.com/10.1097/EDE.0000000000001667},
    doi = {10.1097/EDE.0000000000001667},
    language = {en},
    number = {1},
    urldate = {2025-03-27},
    journal = {Epidemiology},
    author = {Didelez, Vanessa and Haug, Ulrike and Garcia-Albeniz, Xabier},
    month = jan,
    year = {2024},
    pages = {e3--e3},
}

@article{komura2025_exploring,
    title = {Exploring the {Application} of {Target} {Trial} {Emulation} in {Vaccine} {Evaluation}: {Scoping} {Review}},
    issn = {0002-9262},
    shorttitle = {Exploring the {Application} of {Target} {Trial} {Emulation} in {Vaccine} {Evaluation}},
    url = {https://doi.org/10.1093/aje/kwaf053},
    doi = {10.1093/aje/kwaf053},
    urldate = {2025-03-26},
    journal = {American Journal of Epidemiology},
    author = {Komura, Toshiaki and Watanabe, Miwa and Shioda, Kayoko},
    month = mar,
    year = {2025},
    pages = {kwaf053},
}

@article{mansournia2013_matcheda,
    title = {Matched designs and causal diagrams},
    volume = {42},
    issn = {0300-5771},
    url = {https://doi.org/10.1093/ije/dyt083},
    doi = {10.1093/ije/dyt083},
    number = {3},
    urldate = {2025-01-20},
    journal = {International Journal of Epidemiology},
    author = {Mansournia, Mohammad A and Hernán, Miguel A and Greenland, Sander},
    month = jun,
    year = {2013},
    pages = {860--869},
}

@article{sjolander2013_ignoring,
    title = {Ignoring the matching variables in cohort studies – when is it valid and why?},
    volume = {32},
    copyright = {Copyright © 2013 John Wiley \& Sons, Ltd.},
    issn = {1097-0258},
    url = {https://onlinelibrary.wiley.com/doi/abs/10.1002/sim.5879},
    doi = {10.1002/sim.5879},
    language = {en},
    number = {27},
    urldate = {2025-01-20},
    journal = {Statistics in Medicine},
    author = {Sjölander, Arvid and Greenland, Sander},
    year = {2013},
    pages = {4696--4708},
}

@article{andrillon2020_performance,
    title = {Performance of propensity score matching to estimate causal effects in small samples},
    volume = {29},
    issn = {0962-2802},
    url = {https://doi.org/10.1177/0962280219887196},
    doi = {10.1177/0962280219887196},
    language = {EN},
    number = {3},
    urldate = {2025-04-01},
    journal = {Statistical Methods in Medical Research},
    author = {Andrillon, Anais and Pirracchio, Romain and Chevret, Sylvie},
    month = mar,
    year = {2020},
    publisher = {SAGE Publications Ltd STM},
    pages = {644--658},
}

@article{monteiro2023_impact,
    title = {Impact of {CoronaVac} on {Covid}-19 outcomes of elderly adults in a large and socially unequal {Brazilian} city: {A} target trial emulation study},
    volume = {41},
    issn = {0264-410X},
    shorttitle = {Impact of {CoronaVac} on {Covid}-19 outcomes of elderly adults in a large and socially unequal {Brazilian} city},
    url = {https://www.sciencedirect.com/science/article/pii/S0264410X23009088},
    doi = {10.1016/j.vaccine.2023.07.065},
    number = {39},
    urldate = {2025-03-30},
    journal = {Vaccine},
    author = {Monteiro, Higor S. and Lima Neto, Antonio S. and Kahn, Rebecca and Sousa, Geziel S. and Carmona, Humberto A. and Andrade, José S. and Castro, Marcia C.},
    month = sep,
    year = {2023},
    pages = {5742--5751},
}

@book{hernan2020_causal,
  title={Causal Inference: What If},
  author={Hernán , M.A. and Robins, J.M.},
  isbn={9781420076165},
  lccn={2022050839},
  url={https://books.google.com/books?id=_KnHIAAACAAJ},
  year={2020},
  publisher={Boca Raton: Chapman \& Hall/CRC}
}

@article{hulme2023_challenges,
    title = {Challenges in {Estimating} the {Effectiveness} of {COVID}-19 {Vaccination} {Using} {Observational} {Data}},
    volume = {176},
    issn = {0003-4819},
    url = {https://www-acpjournals-org.proxy.library.emory.edu/doi/10.7326/M21-4269},
    doi = {10.7326/M21-4269},
    number = {5},
    urldate = {2024-04-19},
    journal = {Annals of Internal Medicine},
    author = {Hulme, William J. and Williamson, Elizabeth and Horne, Elsie M.F. and Green, Amelia and McDonald, Helen I. and Walker, Alex J. and Curtis, Helen J. and Morton, Caroline E. and MacKenna, Brian and Croker, Richard and Mehrkar, Amir and Bacon, Seb and Evans, David and Inglesby, Peter and Davy, Simon and Bhaskaran, Krishnan and Schultze, Anna and Rentsch, Christopher T. and Tomlinson, Laurie and Douglas, Ian J. and Evans, Stephen J.W. and Smeeth, Liam and Palmer, Tom and Goldacre, Ben and Hernán, Miguel A. and Sterne, Jonathan A.C.},
    month = may,
    year = {2023},
publisher = {American College of Physicians},
    pages = {685--693},
}

@article{meah2023_design,
    title={Design and analysis heterogeneity in observational studies of COVID-19 booster effectiveness: A review and case study},
   author={Meah, Sabir and Shi, Xu and Fritsche, Lars G and Salvatore, Maxwell and Wagner, Abram and Martin, Emily T and Mukherjee, Bhramar},
  journal={Science Advances},
  volume={9},
  number={51},
  pages={eadj3747},
  year={2023},
  publisher={American Association for the Advancement of Science}

}

@article{fung2024_sources,
    title = {Sources of bias in observational studies of covid‐19 vaccine effectiveness},
    volume = {30},
    issn = {1356-1294, 1365-2753},
    url = {https://onlinelibrary.wiley.com/doi/10.1111/jep.13839},
    doi = {10.1111/jep.13839},
    language = {en},
    number = {1},
    urldate = {2025-04-02},
    journal = {Journal of Evaluation in Clinical Practice},
    author = {Fung, Kaiser and Jones, Mark and Doshi, Peter},
    month = feb,
    year = {2024},
    pages = {30--36},
}

@misc{demonte2024_assessing,
    title = {Assessing {COVID}-19 {Vaccine} {Effectiveness} in {Observational} {Studies} via {Nested} {Trial} {Emulation}},
    url = {http://arxiv.org/abs/2403.18115},
    urldate = {2024-03-30},
    publisher = {arXiv},
    author = {DeMonte, Justin B. and Shook-Sa, Bonnie E. and Hudgens, Michael G.},
    month = mar,
    year = {2024},
    note = {arXiv:2403.18115},
}

@article{hernan2008_observational,
    title = {Observational {Studies} {Analyzed} {Like} {Randomized} {Experiments}: {An} {Application} to {Postmenopausal} {Hormone} {Therapy} and {Coronary} {Heart} {Disease}},
    volume = {19},
    issn = {1044-3983},
    shorttitle = {Observational {Studies} {Analyzed} {Like} {Randomized} {Experiments}},
    url = {https://journals.lww.com/00001648-200811000-00002},
    doi = {10.1097/EDE.0b013e3181875e61},
    language = {en},
    number = {6},
    urldate = {2024-05-01},
    journal = {Epidemiology},
    author = {Hernán, Miguel A. and Alonso, Alvaro and Logan, Roger and Grodstein, Francine and Michels, Karin B. and Willett, Walter C. and Manson, JoAnn E. and Robins, James M.},
    month = nov,
    year = {2008},
    pages = {766--779},
}

@article{lin2022_effectiveness,
    title = {Effectiveness of {Covid}-19 {Vaccines} over a 9-{Month} {Period} in {North} {Carolina}},
    volume = {386},
    copyright = {http://www.nejmgroup.org/legal/terms-of-use.htm},
    issn = {0028-4793, 1533-4406},
    url = {http://www.nejm.org/doi/10.1056/NEJMoa2117128},
    doi = {10.1056/NEJMoa2117128},
    language = {en},
    number = {10},
    urldate = {2024-09-19},
    journal = {New England Journal of Medicine},
    author = {Lin, Dan-Yu and Gu, Yu and Wheeler, Bradford and Young, Hayley and Holloway, Shannon and Sunny, Shadia-Khan and Moore, Zack and Zeng, Donglin},
    month = mar,
    year = {2022},
    pages = {933--941},
}

@article{vasileiou2021_interim,
    title = {Interim findings from first-dose mass {COVID}-19 vaccination roll-out and {COVID}-19 hospital admissions in {Scotland}: a national prospective cohort study},
    volume = {397},
    issn = {0140-6736, 1474-547X},
    shorttitle = {Interim findings from first-dose mass {COVID}-19 vaccination roll-out and {COVID}-19 hospital admissions in {Scotland}},
    url = {https://www.thelancet.com/journals/lancet/article/PIIS0140-6736(21)00677-2/fulltext?utm},
    doi = {10.1016/S0140-6736(21)00677-2},
    language = {English},
    number = {10285},
    urldate = {2024-06-04},
    journal = {The Lancet},
    author = {Vasileiou, Eleftheria and Simpson, Colin R. and Shi, Ting and Kerr, Steven and Agrawal, Utkarsh and Akbari, Ashley and Bedston, Stuart and Beggs, Jillian and Bradley, Declan and Chuter, Antony and Lusignan, Simon de and Docherty, Annemarie B. and Ford, David and Hobbs, FD Richard and Joy, Mark and Katikireddi, Srinivasa Vittal and Marple, James and McCowan, Colin and McGagh, Dylan and McMenamin, Jim and Moore, Emily and Murray, Josephine LK and Pan, Jiafeng and Ritchie, Lewis and Shah, Syed Ahmar and Stock, Sarah and Torabi, Fatemeh and Tsang, Ruby SM and Wood, Rachael and Woolhouse, Mark and Robertson, Chris and Sheikh, Aziz},
    month = may,
    year = {2021},
    pmid = {33901420},
    publisher = {Elsevier},
    pages = {1646--1657},
}

@article{keogh2023_causal,
    title = {Causal inference in survival analysis using longitudinal observational data: {Sequential} trials and marginal structural models},
    volume = {42},
    issn = {1097-0258},
    shorttitle = {Causal inference in survival analysis using longitudinal observational data},
    url = {https://onlinelibrary.wiley.com/doi/abs/10.1002/sim.9718},
    doi = {10.1002/sim.9718},
    language = {en},
    number = {13},
    urldate = {2024-03-25},
    journal = {Statistics in Medicine},
    author = {Keogh, Ruth H. and Gran, Jon Michael and Seaman, Shaun R. and Davies, Gwyneth and Vansteelandt, Stijn},
    year = {2023},
    pages = {2191--2225},
}

@article{iacus2012_causal,
    title = {Causal {Inference} without {Balance} {Checking}: {Coarsened} {Exact} {Matching}},
    volume = {20},
    copyright = {https://www.cambridge.org/core/terms},
    issn = {1047-1987, 1476-4989},
    shorttitle = {Causal {Inference} without {Balance} {Checking}},
    url = {https://www.cambridge.org/core/product/identifier/S1047198700012985/type/journal_article},
    doi = {10.1093/pan/mpr013},
    language = {en},
    number = {1},
    urldate = {2025-03-27},
    journal = {Political Analysis},
    author = {Iacus, Stefano M. and King, Gary and Porro, Giuseppe},
    year = {2012},
    pages = {1--24},
}

@article{ho2007_matching,
    title = {Matching as {Nonparametric} {Preprocessing} for {Reducing} {Model} {Dependence} in {Parametric} {Causal} {Inference}},
    volume = {15},
    copyright = {https://www.cambridge.org/core/terms},
    issn = {1047-1987, 1476-4989},
    url = {https://www.cambridge.org/core/product/identifier/S1047198700006483/type/journal_article},
    doi = {10.1093/pan/mpl013},
    language = {en},
    number = {3},
    urldate = {2025-03-27},
    journal = {Political Analysis},
    author = {Ho, Daniel E. and Imai, Kosuke and King, Gary and Stuart, Elizabeth A.},
    year = {2007},
    pages = {199--236},
}

@article{ioannou2022_covid19,
    title = {{COVID}-19 {Vaccination} {Effectiveness} {Against} {Infection} or {Death} in a {National} {U}.{S}. {Health} {Care} {System}},
    volume = {175},
    issn = {0003-4819},
    url = {https://www.acpjournals.org/doi/10.7326/M21-3256},
    doi = {10.7326/M21-3256},
    number = {3},
    urldate = {2024-04-29},
    journal = {Annals of Internal Medicine},
    author = {Ioannou, George N. and Locke, Emily R. and O’Hare, Ann M. and Bohnert, Amy S.B. and Boyko, Edward J. and Hynes, Denise M. and Berry, Kristin},
    month = mar,
    year = {2022},
    publisher = {American College of Physicians},
    pages = {352--361},
}

@article{reis2021_effectiveness,
    title = {Effectiveness of {BNT162b2} {Vaccine} against {Delta} {Variant} in {Adolescents}},
    volume = {385},
    issn = {0028-4793},
    url = {https://www.nejm.org/doi/full/10.1056/NEJMc2114290},
    doi = {10.1056/NEJMc2114290},
    number = {22},
    urldate = {2025-03-30},
    journal = {New England Journal of Medicine},
    author = {Reis, Ben Y. and Barda, Noam and Leshchinsky, Michael and Kepten, Eldad and Hernán, Miguel A. and Lipsitch, Marc and Dagan, Noa and Balicer, Ran D.},
    month = nov,
    year = {2021},
    publisher = {Massachusetts Medical Society}, 
    pages = {2101--2103},
}

@article{hernan2010_hazards,
    title = {The {Hazards} of {Hazard} {Ratios}},
    volume = {21},
    issn = {1044-3983},
    url = {https://journals.lww.com/epidem/fulltext/2010/01000/the_hazards_of_hazard_ratios.4.aspx},
    doi = {10.1097/EDE.0b013e3181c1ea43},
    language = {en-US},
    number = {1},
    urldate = {2024-04-11},
    journal = {Epidemiology},
    author = {Hernán, Miguel A.},
    month = jan,
    year = {2010},
    pages = {13},
}

@article{mcconeghy2022_infections,
    title = {Infections, {Hospitalizations}, and {Deaths} {Among} {US} {Nursing} {Home} {Residents} {With} vs {Without} a {SARS}-{CoV}-2 {Vaccine} {Booster}},
    volume = {5},
    issn = {2574-3805},
    url = {https://doi.org/10.1001/jamanetworkopen.2022.45417},
    doi = {10.1001/jamanetworkopen.2022.45417},
    number = {12},
    urldate = {2025-03-30},
    journal = {JAMA Network Open},
    author = {McConeghy, Kevin W. and Bardenheier, Barbara and Huang, Andrew W. and White, Elizabeth M. and Feifer, Richard A. and Blackman, Carolyn and Santostefano, Christopher M. and Lee, Yoojin and DeVone, Frank and Halladay, Christopher W. and Rudolph, James L. and Zullo, Andrew R. and Mor, Vincent and Gravenstein, Stefan},
    month = dec,
    year = {2022},
    pages = {e2245417},
}

@article{polinski2022_durability,
    title = {Durability of the {Single}-{Dose} {Ad26}.{COV2}.{S} {Vaccine} in the {Prevention} of {COVID}-19 {Infections} and {Hospitalizations} in the {US} {Before} and {During} the {Delta} {Variant} {Surge}},
    volume = {5},
    issn = {2574-3805},
    url = {https://doi.org/10.1001/jamanetworkopen.2022.2959},
    doi = {10.1001/jamanetworkopen.2022.2959},
    number = {3},
    urldate = {2025-04-02},
    journal = {JAMA Network Open},
    author = {Polinski, Jennifer M. and Weckstein, Andrew R. and Batech, Michael and Kabelac, Carly and Kamath, Tripthi and Harvey, Raymond and Jain, Sid and Rassen, Jeremy A. and Khan, Najat and Schneeweiss, Sebastian},
    month = mar,
    year = {2022},
    pages = {e222959},
}

@article{ioannou2022_effectiveness,
    title = {Effectiveness of {mRNA} {COVID}-19 {Vaccine} {Boosters} {Against} {Infection}, {Hospitalization}, and {Death}: {A} {Target} {Trial} {Emulation} in the {Omicron} ({B}.1.1.529) {Variant} {Era}},
    volume = {175},
    issn = {0003-4819},
    shorttitle = {Effectiveness of {mRNA} {COVID}-19 {Vaccine} {Boosters} {Against} {Infection}, {Hospitalization}, and {Death}},
    url = {https://www.acpjournals.org/doi/full/10.7326/M22-1856},
    doi = {10.7326/M22-1856},
    number = {12},
    urldate = {2025-03-30},
    journal = {Annals of Internal Medicine},
    author = {Ioannou, George N. and Bohnert, Amy S.B. and O’Hare, Ann M. and Boyko, Edward J. and Maciejewski, Matthew L. and Smith, Valerie A. and Bowling, C. Barrett and Viglianti, Elizabeth and Iwashyna, Theodore J. and Hynes, Denise M. and Berry, Kristin and {COVID-19 Observational Research Collaboratory (CORC)}},
    month = dec,
    year = {2022},
publisher = {American College of Physicians},
    pages = {1693--1706},
}

@article{gazit_incidence_2022,
    title = {The {Incidence} of {SARS}-{CoV}-2 {Reinfection} in {Persons} {With} {Naturally} {Acquired} {Immunity} {With} and {Without} {Subsequent} {Receipt} of a {Single} {Dose} of {BNT162b2} {Vaccine}},
    volume = {175},
    issn = {0003-4819},
    url = {https://www.acpjournals.org/doi/full/10.7326/M21-4130},
    doi = {10.7326/M21-4130},
    number = {5},
    urldate = {2025-03-30},
    journal = {Annals of Internal Medicine},
    author = {Gazit, Sivan and Shlezinger, Roei and Perez, Galit and Lotan, Roni and Peretz, Asaf and Ben-Tov, Amir and Herzel, Esma and Alapi, Hillel and Cohen, Dani and Muhsen, Khitam and Chodick, Gabriel and Patalon, Tal},
    month = may,
    year = {2022},
publisher = {American College of Physicians},
    pages = {674--681},
}

@misc{harton2025_estimating,
    address = {Rochester, NY},
    type = {{SSRN} {Scholarly} {Paper}},
    title = {Estimating {Covid}-19 {Vaccine} {Effectiveness} {Among} {Children} and {Adolescents} {Using} {Data} from a {School}-{Based} {Weekly} {Covid}-19 {Testing} {Program}},
    url = {https://papers.ssrn.com/abstract=5173302},
    doi = {10.2139/ssrn.5173302},
    language = {en},
    urldate = {2025-04-11},
    author = {Harton, Paige and Chamberlain, Allison and Moore, Amy and Fletcher, Grace and Nelson, Kristin N. and Dean, Natalie and Lopman, Ben and Rogawski McQuade, Elizabeth T.},
    month = mar,
    year = {2025},
    note  = {Preprint},
}

\end{document}

\maketitle 
\tableofcontents 

\newpage
%\onehalfspacing
\addcontentsline{toc}{section}{eAppendix 1- Identification}
\section*{eAppendix 1- Identification}\label{sec:identification_proof}

\subsection*{Proof of Proposition 1:}

We assume the ordering $(\tilde{C}_k, \tilde{Y}_k, V_k)$, where these variables are defined in the main text. We let $Y_k(d,v) = I(Y(d,v) \leq k)$ denote the counterfactual indicator of observing an endpoint by day $k$.  We note that SUTVA implies the consistency assumption which can be stated as 
\[\text{If }  \overline{V}_{k-1} = \bar{\nu}_{k-1}^{d,v} \text{ and } C_{k} = 0, \text{ then } \bar{Y}_k(d,v) = \bar{\tilde{Y}}_k  .\]
This assumption states that if an individual's observed vaccination history through day $k -1$ and censoring history through $k$ is the same as the vaccination and censoring history that would be assigned by the intervention of interest through the same timepoints, then the counterfactual event history under the intervention of interest through day $k$ equals the observed event history through day $k$. Note that this also implies that the counterfactual event histories $\bar{Y}_k(d,v)$ for different vaccine interventions can be the same if the interventions are the same up to day $k$.

\begin{proof}
Assumption 3 implies that \[
\psi_v(t_0; d,x) = P[Y(d,v) \leq d + t_0 \mid Y(d,v) > d + \tau, X = x]  \]
where one conditioning set associated with the principal stratum is dropped. 

By simple rules of probability,
\begin{align}
\psi_v(t_0; d,x) &= 1 -  P[Y(d,v) > d + t_0 \mid Y(d,v) > d + \tau, X = x] \notag \\
&= 1 -  \frac{P[Y(d,v) > d + t_0 \mid  X = x]}{P[Y(d,v) > d + \tau \mid X = x]} \notag \\
& = 1 -  \frac{P[Y_{d + t_0}(d,v) = 0 \mid  X = x]}{P[Y_{d + \tau}(d,v) = 0 \mid X = x]} \label{eqn:psi} \ .
\end{align}

This quantity is identified if the survival probabilities in the numerator and denominator are identified. We show below that the survival probability at a general time $t$ is indeed identifiable. 

First, note that the survival probability at time $t$ can be equivalently written as 
\begin{align*}
 P[Y_t(d,v) =0 \mid  X = x]  &=  P[Y_t(d,v) =0 \mid  X = x, \overline{V}_{-1} = \bar{\nu}_{-1}^{d,v}, \tilde{C}_0 = 0, \tilde{Y}_0 = 0]  \\
 &=  P[Y_t(d,v) =0 \mid  X = x, \overline{V}_{0} = \bar{\nu}_{0}^{d,v}, \tilde{C}_1 = 0, \tilde{Y}_0 = 0] \, 
\end{align*}

where the first equality is due to the  definition of the study population and the second equality is due to exchangeability of vaccination and censoring. 

Applying the law of total probability and then consistency (which follows from the stable unit treatment value assumption), we obtain 

\small{
\begin{align}
& \sum_{n =0}^1 P[Y_t(d,v) =0 \mid  X = x, \overline{V}_{0} = \bar{\nu}_{0}^{d,v}, \tilde{C}_1 = 0, \tilde{Y}_0 = 0, Y_1(d,v) = n]   P[Y_1(d,v) = n  \mid  X = x, \overline{V}_{0} = \bar{\nu}_{0}^{d,v}, \tilde{C}_1 = 0, \tilde{Y_0} = 0] \notag \\
& = \sum_{n =0}^1 P[Y_t(d,v) =0 \mid  X = x, \overline{V}_{0} = \bar{\nu}_{0}^{d,v}, \tilde{C}_1 = 0, \tilde{Y}_0 = 0, Y_1(d,v) = n]  P[\tilde{Y_1} = n  \mid  X = x, \overline{V}_{0} = \bar{\nu}_{0}^{d,v}, \tilde{C}_1 = 0, \tilde{Y_0} = 0] \ . \
\end{align}
}

Using the fact that $P[Y_t(d,v) = 0 \mid Y_1(d,v) = 1] = 0$, the sum simplifies to  the case when $n = 0$. Applying consistency, this simplifies to 
\begin{equation}
\begin{aligned}
&  P[Y_t(d,v) =0 \mid  X = x, \overline{V}_{0} = \bar{\nu}_{0}^{d,v}, \tilde{C}_1 = 0, \tilde{Y}_1 = 0] \\
& \hspace{4em} \times P[\tilde{Y}_1 = 0  \mid  X = x, \overline{V}_{0} = \bar{\nu}_{0}^{d,v}, \tilde{C}_1 = 0, \tilde{Y}_0 = 0] \ .
\end{aligned} 
\end{equation}
\\
Then, applying sequential exchangeability of vaccination and censoring to the first term in the product, we can write
\begin{equation}
\begin{aligned}
& \hspace{2em}  P[Y_t(d,v) =0 \mid  X = x, \overline{V}_{1} = \bar{\nu}_{1}^{d,v}, \tilde{C}_2 = 0, \tilde{Y}_1 = 0] \\
& \hspace{4em} \times P[\tilde{Y}_1 = 0  \mid  X = x, \overline{V}_{0} = \bar{\nu}_{0}^{d,v}, \tilde{C}_1 = 0,\tilde{Y}_0 = 0]
\end{aligned}   
\end{equation}

Repeatedly applying the law of total probability, consistency, and sequential exchangeability,  as in equations (2)-(4),  to terms like $ P[Y_t(d,v) =0 \mid  X = x, \overline{V}_{1} = \bar{\nu}_{1}^{d,v}, \tilde{C}_2 = 0, \tilde{Y}_1 = 0]$, we eventually obtain the result
\begin{align*}
& P[Y_t(d,v) =0 \mid  X = x] \\
& = \prod_{s = 1}^t P[\tilde{Y}_s = 0  \mid  X = x, \overline{V}_{s -1} = \bar{\nu}_{s-1}^{d,v}, \tilde{C}_s = 0, \tilde{Y}_{s-1} = 0] \\
& = \prod_{s = 1}^t \{1 - P[\tilde{Y}_s = 1  \mid  X = x, \overline{V}_{s -1} = \bar{\nu}_{s-1}^{d,v}, \tilde{C}_s = 0, \tilde{Y}_{s-1} = 0]\} \ .
\end{align*}

Substituting this result into Equation (\ref{eqn:psi}), we find that the general identification form for $\psi_v(t_0; d,x)$ is
\begin{align*}
\psi_v(t_0; d,x)  &= 1 - \frac{\prod_{s = 1}^{d + t_0} \{1 - P[\tilde{Y}_s = 1  \mid  X = x, \overline{V}_{s -1} = \bar{\nu}_{s-1}^{d,v}, \tilde{C}_s = 0, \tilde{Y}_{s-1} = 0]}{\prod_{s = 1}^{d + \tau} \{1 - P[\tilde{Y}_s = 1  \mid  X = x, \overline{V}_{s -1} = \bar{\nu}_{s-1}^{d,v}, \tilde{C}_s = 0, \tilde{Y}_{s-1} = 0]} \\
&= 1 - \prod_{s = d + \tau + 1}^{d + t_0} \{1 - P[\tilde{Y}_s = 1  \mid  X = x, \overline{V}_{s -1} = \bar{\nu}_{s-1}^{d,v}, \tilde{C}_s = 0, \tilde{Y}_{s-1} = 0] \} \ .
\end{align*}

This identification can be further simplified when we consider specific values for $v$. 

For $v = 0$, Assumption 4 implies that $\bar{\nu}_{s-1}^{d,0} = \bm{0}_{s-1}$ for all $d$. Thus, we can  write 
\begin{align*}
\psi_0(t_0; d,x) &=  1 - \prod_{s = d + \tau + 1}^{d + t_0} \{1 - P[\tilde{Y}_s = 1  \mid  X = x, \bar{V}_{s-1} = \bm{0}_{s-1}, \tilde{C}_s = 0, \tilde{Y}_{s-1} = 0]\}\\
& =  1 - \prod_{s = d + \tau + 1}^{d + t_0} \{1 - P[\tilde{Y} = s \mid  X = x, \bar{V}_{s-1} = \bm{0}_{s-1}, \tilde{C}_s = 0, \tilde{Y} >  s - 1]\}
\end{align*}

For $v = 1$, we note that whenever $s > d$, the set of individuals with $\overline{V}_{s-1} = \bar{\nu}_{s-1}^{d,1} = (\bm{0}_{d-1}, 1, \bm{0}_{s-1 - d})$ is equivalent to the set of individuals with  $D^* = d$. Thus, we can write
\begin{align*}
\psi_1(t_0; d,x) &=  1 - \prod_{s = d + \tau + 1}^{d + t_0} \{1 - P[\tilde{Y}_s = 1  \mid  X = x, D^* = d, \tilde{C}_s = 0, \tilde{Y}_{s-1} = 0]\} \\
&=  1 - \prod_{s = d + \tau + 1}^{d + t_0} \{1 - P[\tilde{Y} = s  \mid  X = x, D^* = d, \tilde{C}_s = 0, \tilde{Y} > s - 1]\}\\
&=  1 - \prod_{j = \tau + 1}^{ t_0} \{1 - P[\tilde{Y} - d = j  \mid  X = x, D^* = d, \tilde{C}_{j + d} = 0, \tilde{Y} - d > j - 1]\} \\
&=  1 - \prod_{j = \tau + 1}^{d + t_0} \{1 - P[\tilde{T} = j  \mid  X = x, D^* = d,  \tilde{C}_{j + d} = 0, \tilde{T} > j - 1]\} \,
\end{align*}

where $\tilde{T} = \tilde{Y} - D^*$.

\end{proof}

\addcontentsline{toc}{section}{eAppendix 2- Bootstrapped confidence intervals}
\section*{eAppendix 2- Bootstrapped confidence intervals}

In this section, we describe the construction of pointwise and simultaneous confidence intervals for cumulative incidence and VE estimates. It is  explicitly written using notation for the proposed parameters, but the same procedures were used for the matching-based parameters. 

\subsection*{Pointwise confidence intervals}
Wald-style confidence intervals based on bootstrapped standard errors were used. Confidence intervals for $\hat{\bar{\psi}}_v(t_0)$ were constructed by forming confidence intervals on the $\text{logit}(x) = \log(\frac{x}{1-x})$  scale and then converting the resulting confidence limits back to the cumulative incidence scale via the inverse-logit transformation  $\text{logit}^{-1}(x) = \frac{e^x}{1 + e^x}$.  The $100(1 - \alpha)$\%  confidence limits for  $\hat{\bar{\psi}}_v(t_0)$  were thus given by \[\text{expit}\left\{\text{logit}\{\hat{\bar{\psi}}_v(t_0)\} \pm z_{1 - \alpha/2} \ \text{SE}_{boot}\big(\text{logit}\{\hat{\bar{\psi}}_v(t_0)\}\big)\right\} \ , \]
where $z_{1-\alpha/2}$ is the $(1 - \alpha/2)$-quantile of a standard Normal random variable.   

\noindent Confidence intervals for $\widehat{\text{VE}}_{\text{C}}(t_0)$ were constructed by forming confidence intervals on the $\log(1 - x)$ scale and converting the resulting confidence limits back to the VE scale via the transformation $1 - \exp(x)$.  The  $100(1 - \alpha)$\% confidence limits for $\widehat{\text{VE}}_{\text{C}}(t_0)$ were thus given by  \[1 - \text{exp}\left\{\text{log}\{1 - \widehat{\text{VE}}_{\text{C}}(t_0)\} \pm z_{1 - \alpha/2} \ \text{SE}_{boot}\big(\text{log}\{1 - \widehat{\text{VE}}_{\text{C}}(t_0)\}\big)\right\} \ . \]

\subsection*{Simultaneous confidence intervals}

The pointwise confidence intervals above provide a range of uncertainty around an estimate at a single timepoint. 
%such that the confidence interval contains the true value with $(1 - \alpha)$\% confidence. 
In contrast, simultaneous confidence intervals can provide a range of uncertainty around a curve of estimates over a range of timepoints, say $t_1, \ldots, t_M$. 
%such that the confidence band contains the true curve with $(1 - \alpha)$\% confidence. 

$(100 - \alpha)$\% simultaneous confidence intervals for $\bar{\psi}_v(t 
)$ and $\text{VE}_{\text{C}}(t)$ for $t \in \{t_1, \ldots, t_M\}$ can be computed as

\[\text{expit}\left\{\text{logit}\{\hat{\bar{\psi}}_v(t)\} \pm m_{1-\alpha} \ \text{SE}_{boot}\big(\text{logit}\{\hat{\bar{\psi}}_v(t 
)\}\big)\right\} \ , \]

and  

\[1 - \text{exp}\left\{\text{log}\{1 - \widehat{\text{VE}}_{\text{C}}(t)\} \pm m_{1-\alpha} \ \text{SE}_{boot}\big(\text{log}\{1 - \widehat{\text{VE}}_{\text{C}}(t)\}\big)\right\} \ , \]
 where $m_{1 - \alpha}$ is the $(1 - \alpha)$ quantile of the random variable $\max_{t_1 \leq t \leq t_M}|\frac{W_t}{SE(W_t)}|$ where $\bm{W}$ is a multivariarate normal random variable with mean $\bm{0}$ and covariance matrix equal to $\bm{\Sigma} = \text{Cov}(\hat{f}_1,..., \hat{f}_M)$ where $\hat{f}_1,..,\hat{f}_M$ are the appropriate estimates at time points $t_1,...,t_M$ respectively. In practice, $\bm{\Sigma}$ can be estimated by taking the empirical covariance of a matrix of bootstrapped estimates where the rows represent bootstrap iterations and the columns represent each timepoint. The quantile $m_{1-\alpha}$ can be approximated by simulating values of  $\max_{t_1 \leq t \leq t_M}|\frac{W_t}{SE(W_t)}|$ 
 a large number of times, say $N=10,000$ \autocite{ruppert2003_inference}.
 
\addcontentsline{toc}{section}{eTable 1- Simulation design and parameters}
\section*{eTable 1 -  Simulation design and parameters}

The simulation design is summarized in Table \ref{tab:simulation_design}. Covariates, vaccination status, and censoring times were simulated from the probability distributions described. Indicators for day of first vaccination ($D_k$) with placebo or active vaccine were simulated using a logistic model with a probability that depended on covariates and calendar time. Indicators for the study endpoint ($Y_k$) were simulated by first simulating exposure times and then simulating endpoints given exposure using a logistic model that depended on covariates and calendar time.

\begin{table}[t]
    \vspace{-1cm}
    \caption{Summary of simulation design}
    \hspace{-1cm}
    \begin{tabular}{lll}
    \hline 
    Variable & Data generating process and parameters\\
    \hline 
    Covariates & $X_1 \text{ (``male sex") } \sim \text{Binom}(0.5)$\\
     & $X_2 \text{ (``age") } \sim \text{Discrete Uniform}(5,11)$\\
     & $X_3 \text{ (``race”) } \sim$ Multinomial(White = .27, Black = .58, Other = .15) \\ 
     & $X_4 \text{ (``school cluster”)} \sim \text{Discrete Uniform}(1,2,...,10)$ \vspace{2mm} \\
    Vaccination & $V  \sim \text{Binom}(0.42)$ \vspace{2mm} \\
     Censoring & $C \sim \begin{cases}
    \text{Unif}(1,210) & \text{ with probability 10\%}\\
    90 & \text{ with probability 10\%}\\
    210 & \text{ with probability 80\%}
    \end{cases}$   \vspace{2mm} \\ 
    \hline \\ 
    Time to vaccination & $\text{logit Pr}(D_k = 1 \mid X) = \gamma_0(k) + \gamma_XX$, where   \\
    & \; $\gamma_0(k) = .067k \ \text{I}( k \leq 15) + \min(0.1 + 10^{-6}(k-15)^2, 1) $ \\  
    & \; $\gamma_{X_1, male} = \log(1.02) $      \\
    & \; $\gamma_{X_{2}(age)} = \log(0.99)$      \\
    & \; $\gamma_{X_3, black} = \log(0.30)$   \\
    & \; $\gamma_{X_3, other} = \log(0.75)$     \\ 
    & \; $\gamma_{X_4, cluster 2} = \log(2)$  \\
    & \; $\gamma_{X_4, cluster 3} = \log(.8)$ \\
    & \; $\gamma_{X_4, cluster 4} = \log(1.65)$  \\
    & \; $\gamma_{X_4, cluster 5} = \log(1.15)$  \\
    & \; $\gamma_{X_4, cluster 6} = \log(2.45)$  \\
    & \; $\gamma_{X_4, cluster 7} = \log(2.4) $ \\
    & \; $\gamma_{X_4, cluster 8} = \log(1.1) $ \\
    & \; $\gamma_{X_4, cluster 9} = \log(1.1) $ \\
    & \; $\gamma_{X_4, cluster 10} = \log(0.95)$  \vspace{2mm} \\
    \hline \\ 
    Time to next exposure at time $k$ & $E_k \sim \text{Poisson}(\eta(k))$, where \\
    & \; $\eta(k) = \max(50 - .01k, 1)$ \vspace{2mm}  \\
    Time to endpoint  & $\text{logit Pr}(Y_k = 1 \mid X, \text{exposed at } k) = \beta_0(k) + \beta_XX + \beta_v(k)V_{k - \tau}$ , where \\
    & \; $\beta_0(k) \in [-Inf, -1.62] \text{ depending on } k$ \\ 
    & \; $\beta_{X_1, male}    = \log(1.1)$\\
    & \; $\beta_{X_{2}(age)}   = \log(.95)$\\
    & \; $\beta_{X_3, black}   = \log(1.15)$\\
    & \; $\beta_{X_3, other}   = \log(.8)  $\\
    & \; $\beta_{X_4, cluster2} = \log(1.1) $ \\
    & \; $\beta_{X_4, cluster3} = \log(0.7) $ \\
    & \; $\beta_{X_4, cluster4} = \log(1.3) $ \\
    & \; $\beta_{X_4, cluster5} = \log(.97) $ \\
    & \; $\beta_{X_4, cluster6} = \log(1.2) $ \\
    & \; $\beta_{X_4, cluster7} = \log(1.8) $  \\
    & \; $\beta_{X_4, cluster8} = \log(.8)  $ \\
    & \; $\beta_{X_4, cluster9} = \log(0.8) $ \\
    & \; $\beta_{X_4, cluster10} = \log(0.85)$ \\ 
    & \; $\beta_v(k) = \log(\min( (2.5*10^{-5})k^2, 1))$  \vspace{2mm} \\
    \hline
    \end{tabular}
    \label{tab:simulation_design}
\end{table}

\clearpage

\addcontentsline{toc}{section}{eTable 2- Simulation results for cumulative incidences}
\section*{eTable 2- Simulation results for cumulative incidences}

\begin{table}[h]
%\vspace{-2cm}
\centering
\caption{Simulation study results for cumulative incidence estimation based on 1000 simulations}
\centering
\begin{threeparttable}
\begin{tabular}[t]{rlccccc}
\toprule
\textbf{N} & \textbf{Method} & \textbf{Bias} & \textbf{MSE} & \textbf{Coverage} & \textbf{Width} & \textbf{Rel.Eff.}\\
\midrule
\addlinespace[0.3em]
\multicolumn{7}{l}{\textbf{Cumulative Incidence: no vaccine}}\\
\hline
\hspace{1em} & matching & 0.0001 & 0.0007 & 0.930 & 1.953 & 1.000\\

\hspace{1em}\multirow[t]{-2}{*}{\raggedleft\arraybackslash 500} & proposed & 0.0005 & 0.0002 & 0.970 & 1.034 & 0.261\\
\cmidrule{1-7}
\hspace{1em} & matching & 0.0013 & 0.0003 & 0.965 & 1.211 & 1.000\\

\hspace{1em}\multirow[t]{-2}{*}{\raggedleft\arraybackslash 1000} & proposed & -0.0002 & 0.0001 & 0.972 & 0.681 & 0.303\\
\cmidrule{1-7}
\hspace{1em} & matching & 0.0002 & 0.0001 & 0.963 & 0.752 & 1.000\\

\hspace{1em}\multirow[t]{-2}{*}{\raggedleft\arraybackslash 2000} & proposed & -0.0001 & 0.0000 & 0.959 & 0.464 & 0.409\\
\cmidrule{1-7}
\hspace{1em} & matching & -0.0002 & 0.0000 & 0.944 & 0.445 & 1.000\\

\hspace{1em}\multirow[t]{-2}{*}{\raggedleft\arraybackslash 5000} & proposed & -0.0001 & 0.0000 & 0.952 & 0.288 & 0.406\\
\cmidrule{1-7}
\addlinespace[0.3em]
\multicolumn{7}{l}{\textbf{Cumulative Incidence:  vaccine}}\\
\hline
\hspace{1em} & matching & -0.0010 & 0.0004 & 0.837 & 2.091 & 1.000\\

\hspace{1em}\multirow[t]{-2}{*}{\raggedleft\arraybackslash 500} & proposed & -0.0004 & 0.0002 & 0.959 & 1.991 & 0.475\\
\cmidrule{1-7}
\hspace{1em} & matching & 0.0002 & 0.0002 & 0.954 & 1.602 & 1.000\\

\hspace{1em}\multirow[t]{-2}{*}{\raggedleft\arraybackslash 1000} & proposed & 0.0002 & 0.0001 & 0.978 & 1.187 & 0.567\\
\cmidrule{1-7}
\hspace{1em} & matching & -0.0002 & 0.0001 & 0.966 & 0.977 & 1.000\\

\hspace{1em}\multirow[t]{-2}{*}{\raggedleft\arraybackslash 2000} & proposed & 0.0000 & 0.0000 & 0.965 & 0.790 & 0.723\\
\cmidrule{1-7}
\hspace{1em} & matching & -0.0003 & 0.0000 & 0.960 & 0.566 & 1.000\\

\hspace{1em}\multirow[t]{-2}{*}{\raggedleft\arraybackslash 5000} & proposed & 0.0000 & 0.0000 & 0.953 & 0.486 & 0.766\\
\bottomrule
\end{tabular}
\begin{tablenotes}
\item MSE = mean squared error; Coverage = coverage of nominal 95\% Wald-style bootstrap confidence intervals; Width = average confidence interval width; Rel.Eff. = ratio of the MSE of proposed estimator vs. matching estimator. Both Width and Rel.Eff. are on the logit scale. The true value of the cumulative incidences  in this scenario were 0.059 and .037 for unvaccinated and vaccinated, respectively.
\end{tablenotes}
\end{threeparttable}
\end{table}

\clearpage 

\addcontentsline{toc}{section}{eFigure 1- Matching estimation and bootstrapping}
\section*{eFigure 1- Matching estimation and bootstrapping}

\begin{figure}[h]
    \centering\includegraphics[width=.9\linewidth]{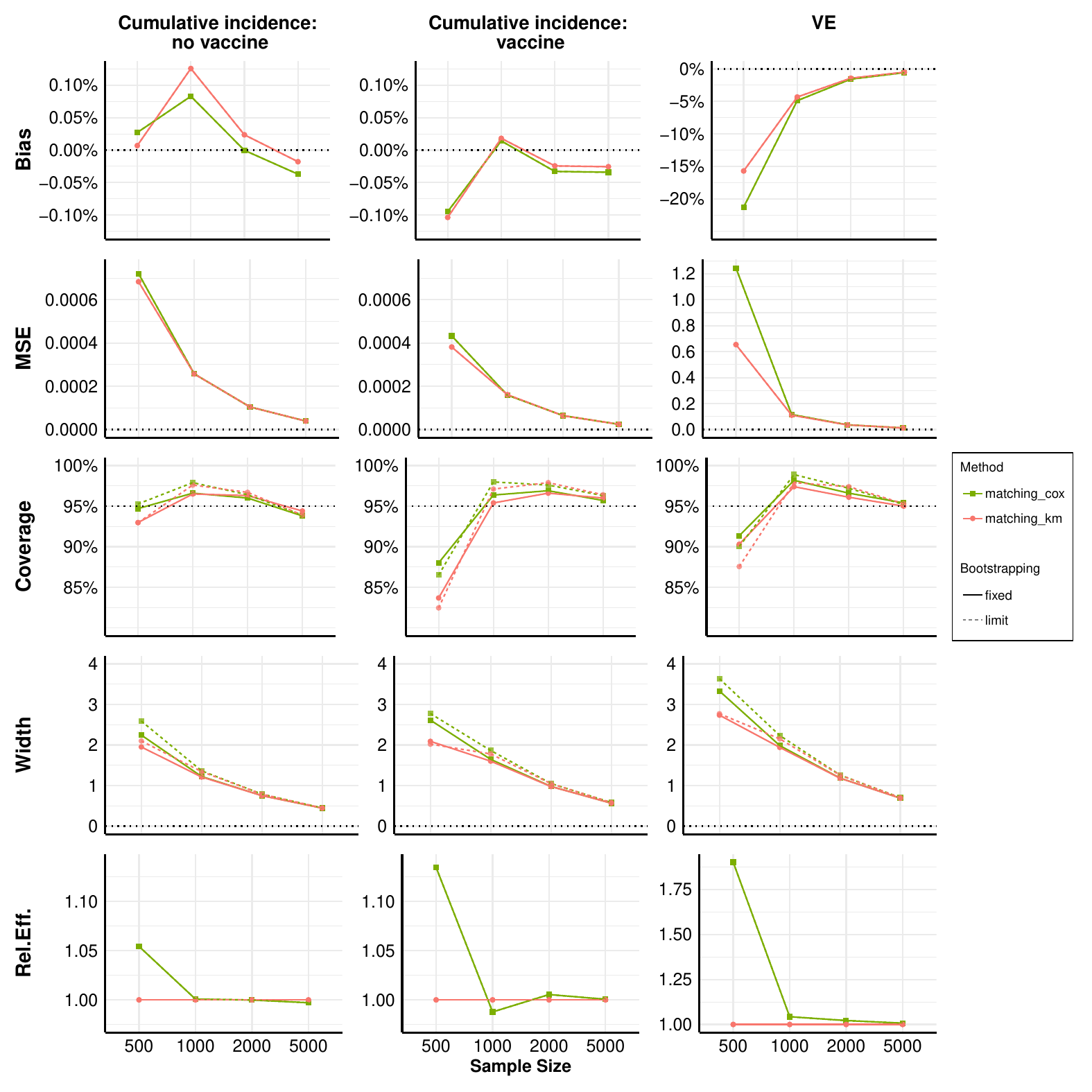}
    \caption{Simulation results for Kaplan Meier vs. Cox regression-based matching estimators. For Cox regression-based estimators, Cox models were fit in the vaccinated and unvaccinated matched groups separately. Method of bootstrapping is expected to affect only coverage and confidence interval width; fixed bootstrapping refers to resampling matched pairs from the same matched dataset whereas limit bootstrapping refers to resampling the observed data and creating new matched datasets each time. }
\end{figure}

\clearpage 

\addcontentsline{toc}{section}{eFigure 2- Variability of matching-estimator with respect to random seed used for matching}
\section*{eFigure 2- Variability of matching-estimator with respect to random seed used for matching }

\begin{figure}[h]
    \centering
    \includegraphics[width=\linewidth]{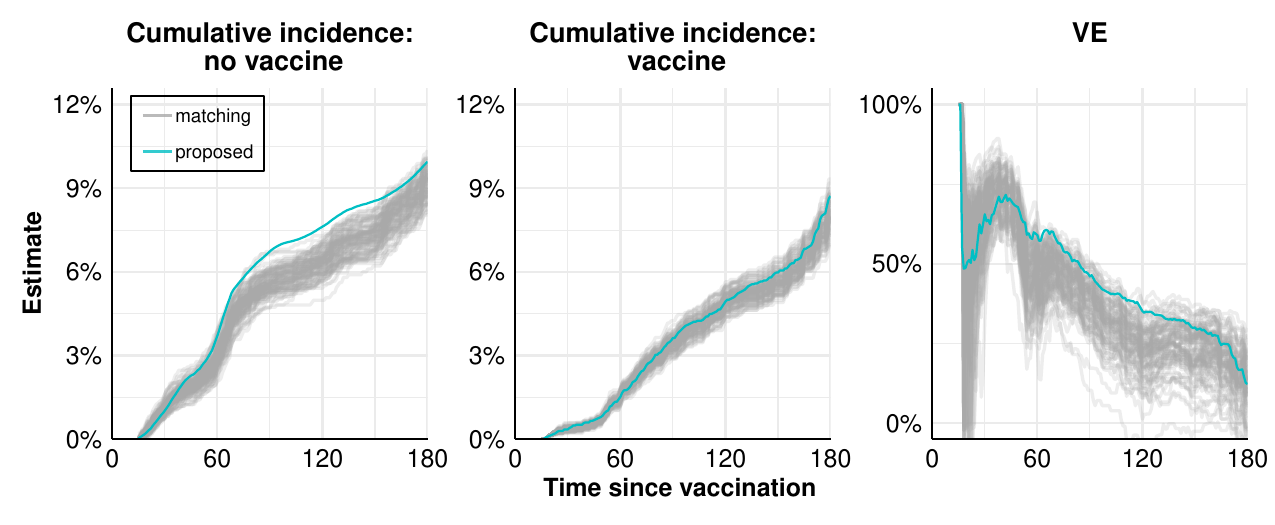}
    \caption{Matching-based estimates using 100 different random seeds in the illustrative study on the effectiveness of the Pfizer-BioNTech COVID-19 vaccine in children 5-11 years old.}
\end{figure}

\clearpage

\addcontentsline{toc}{section}{eFigure 3- Simultaneous confidence intervals for application}
\section*{eFigure 3- Simultaneous confidence intervals for application }

\begin{figure}[h]
    \centering
    \includegraphics[width=1\linewidth]{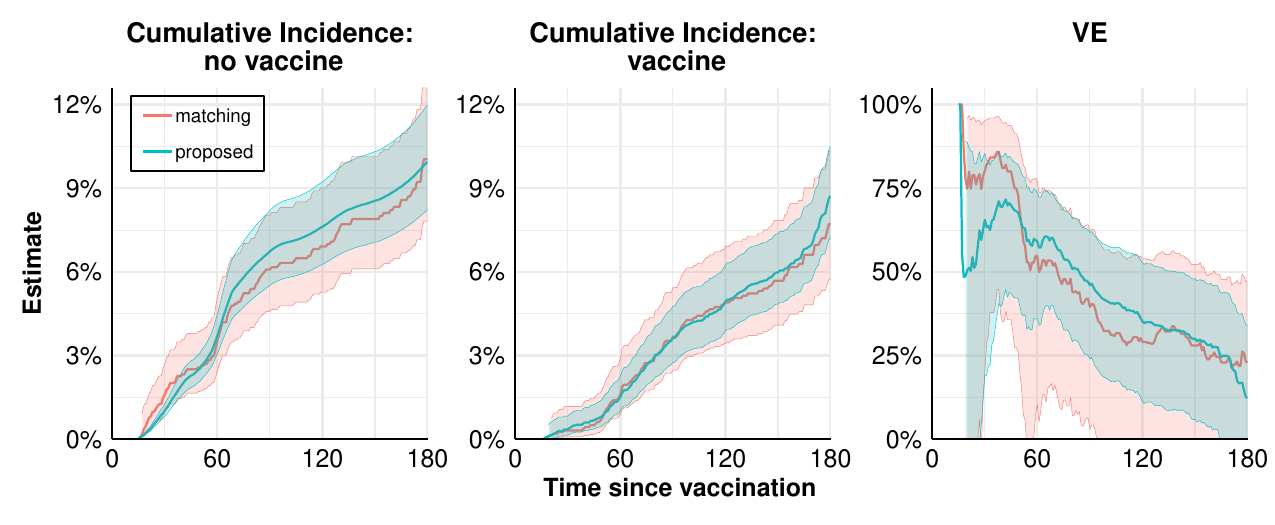}
    \caption{Cumulative incidence of SARS-CoV-2 infection in children 5-11 years old and VE over time. Shaded areas represent 95\% simultaneous Wald-style confidence intervals based on 1000 bootstrap resamples. }
    \label{fig:ve}
\end{figure}

\clearpage 
\printbibliography